\documentclass[a4paper,fleqn,usenatbib]{mnras}

\usepackage[T1]{fontenc}
\usepackage{ae,aecompl}

\usepackage{graphicx}	
\usepackage{amsmath}	
\usepackage{amssymb}	

\graphicspath{{figures/}}  

\newenvironment{tightcenter}{%
	\setlength\topsep{0pt}
	\setlength\parskip{0pt}
	\begin{center}
}{%
 	\end{center}
}

\title[Fluid model of the Centaurus\,A jet]{A 1D fluid model of the Centaurus\,A jet}

\author[S. Wykes et al.]
{Sarka Wykes,$^{1,2}$\thanks{E-mail: swch@protonmail.ch} Bradford T. Snios,$^{2}$ Paul E. J. Nulsen,$^{2,3}$ Ralph P. Kraft,$^{2}$ \newauthor Mark Birkinshaw,$^{4}$ Martin J. Hardcastle,$^{5}$ Diana M. Worrall,$^{4}$ Iain McDonald,$^{6}$ \newauthor Marina Rejkuba,$^{7}$ Thomas W. Jones,$^{8}$  David J. Stark,$^{9}$ William R. Forman,$^{2}$ \newauthor Eileen T. Meyer$^{10}$ and Christine Jones$^{2}$
\\
$^{1}$Independent researcher\\
$^{2}$Harvard-Smithsonian Center for Astrophysics, 60 Garden Street, Cambridge, MA 02138, USA\\
$^{3}$ICRAR, University of Western Australia, 35 Stirling Highway, Crawley, WA 6009, Australia\\
$^{4}$HH Wills Physics Laboratory, University of Bristol, Tyndall Avenue, Bristol BS8 1TL\\
$^{5}$Centre for Astrophysics Research, School of Physics, Astronomy and Mathematics, University of Hertfordshire, College Lane, \\Hatfield, Hertfordshire AL10 9AB\\
$^{6}$Jodrell Bank Centre for Astrophysics, Alan Turing Building, Manchester M13 9PL\\
$^{7}$ESO, Karl-Schwarzschild-Stra\ss e 2, D-85748 Garching, Germany\\
$^{8}$School of Physics and Astronomy and the Minnesota Supercomputing Institute, University of Minnesota, Minneapolis, NM 55455, USA\\
$^{9}$Los Alamos National Laboratory, Los Alamos, NM 87545, USA\\
$^{10}$Department of Physics, University of Maryland Baltimore County, Baltimore, MD 21250, USA\\
}

\date{Accepted 2019 January 31. Received 2019 January 25; in original form 2018 September 27}

\pubyear{2018}

\begin{document}
\label{firstpage}
\pagerange{\pageref{firstpage}--\pageref{lastpage}}
\maketitle

\begin{abstract}
We implement a steady, one-dimensional flow model for the X-ray jet of Centaurus\,A in which entrainment of stellar mass loss is the primary cause of dissipation. Using over 260\,ks of new and archival {\it Chandra}/ACIS data, we have constrained the temperature, density and pressure distributions of gas in the central regions of the host galaxy of Centaurus\,A, and so the pressure throughout the length of its jet. The model is constrained by the observed profiles of pressure and jet width, and conserves matter and energy, enabling us to estimate jet velocities, and hence all the other flow properties. Invoking realistic stellar populations within the jet, we find that the increase in its momentum flux exceeds the net pressure force on the jet unless only about one half of the total stellar mass loss is entrained. For self-consistent models, the bulk speed only falls modestly, from $\sim0.67c$ to $\sim0.52c$ over the range of $0.25-5.94$\,kpc from the nucleus. The sonic Mach number varies between $\sim5.3$ and $3.6$ over this range.
\end{abstract}




\section{Introduction} \label{sec:intro}

Extragalactic radio sources in elliptical galaxies are powered by relatively narrow jets that propagate through the galactic atmospheres of their parent ellipticals. \cite{YOU86} evaluated the momentum transfer by extragalactic jets to the ambient gas and asserted that the momentum transfer of lower-power flows will cause the jets to decelerate, while \cite{BIC94} noted that `FR\,I/BL\,Lac unification requires the initially relativistic jets to have been decelerated somewhere between the parsec and kiloparsec scale'. \cite{BEG82} stressed that a jet can slow down without being completely decollimated but only in the presence of an external galactic pressure gradient. The idea that jets are thermal-pressure confined on kpc scales has been supported by observations of jet geometry (e.g.\,\citealp{CHA80, BRI80}) and the need for an extra confining agent, in addition to magnetic hoop stresses, which cannot function alone (e.g.\,\citealp{EICH82, EICH93, BEG95, KOH12}). There is a good deal of evidence (e.g.\,\citealp{LAI99, LAI02b, CAN04, CAN05, LAI06, KHA12, PER14b, MEY17}) that Fanaroff-Riley class\,I (FR\,I; \citealp{FAN74}) jets decelerate from relativistic to subrelativistic speeds progressively over scales of $\sim0.1-15$\,kpc, the likely cause of this slowing being mass entrainment (e.g.\,\citealp{FAN82, YOU86, BIC94, KOM94, BOW96, LAI02a, LAI02b, HUB06, WYK15a}). 

To allow a jet flow, extragalactic jets must be charge neutral, with electrons and positrons (or electrons and heavier positively-charged particles, or some mixture of these; see e.g.\,\citealp{FAN18}) flowing outwards with similar densities and speeds (e.g.\,\citealp{BEG84}). The velocity and density of jets, and even more so the pressure, are difficult to ascertain rigorously. Little is also known about the element abundances in the material that the jets might acquire, with the work by \cite{WYK15a} predicting an admixture of solar-like composition on kpc scales in the FR\,I source Centaurus\,A, being a notable exception. Jets with a pure electron-positron content are in principle not ruled out by energy and density considerations (e.g.\,\citealp{BIC01, CAR02}), although they may be difficult to keep stable over long distances because of their comparatively low momentum. \cite{CRO05} argued for effectively electron-positron jets in FR\,IIs based on pressure-balance needs,\footnote{An admixture of hadrons, in quantities and energies that do not affect the lobe pressure constraints, is allowed.} and recently \cite{SNI18} have shown that a model including such a jet is tenable in the FR\,II source Cygnus\,A using momentum flux and kinetic power estimates. A large-sample comparison by \cite{CRO18} of FR\,I and FR\,II lobe particle content, inferred from comparison of internal plasma conditions with the external pressure, provides evidence that the two populations are physically different systems with different particle content. 

Classed as FR\,I, Centaurus\,A can be regarded as a misaligned BL\,Lac in the unification scheme (e.g.\,\citealp{CHI01}). It is the nearest radio galaxy at $3.8\pm0.1$\,Mpc \citep{HARR10} -- at which distance 1\,arcsec corresponds to 18.4\,pc -- and is hosted by the elliptical galaxy NGC\,5128. The parent elliptical has a stellar content largely made of two distinct old populations \citep{REJ11}: about $75$ per cent of age about $12$\,Gyr and approximately $25$ per cent of about $3$\,Gyr. The galaxy mass-to-light ratio ($M/L$) is lower than typical for ellipticals (e.g.\,\citealp{HUI95, PEN04a}).

The considerably brighter of the twin jets, referred to as `the jet' in what follows, has attracted observers' attention since the late 1970s. Evident in ultraviolet, optical and infrared images is a prominent dust lane, rich in cold and warm gas and young stars, crossing the central parts of the galaxy (e.g.\,\citealp{DUF79, EBN83, ECK90, QUI06}). The dust lane renders the jet undetectable at optical to ultraviolet wavelengths over its inner $\sim1$\,kpc, and it also leads to contamination of surface brightness profiles at the wavelengths from far-infrared to X-ray over that region, making the kind of studies attempted in this work challenging.
\begin{figure}
	\begin{tightcenter}
	\includegraphics[width=0.48\textwidth]{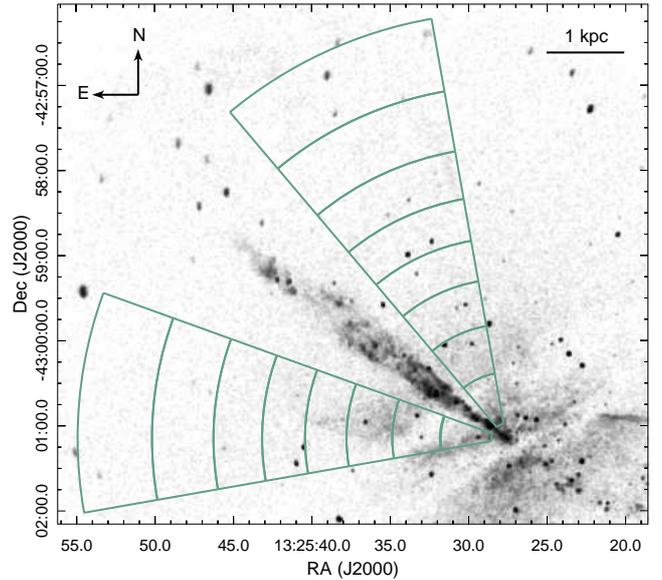}
	\end{tightcenter}
        \caption{Background-subtracted, exposure-corrected {\it Chandra} image in the 0.6--2.0\,keV energy range of Centaurus\,A's jet and its surroundings. Superposed are regions used in the spectral deprojection method (described in Section\,\ref{sec:deprojection}), for which the point sources were excised. Both eastern and north-eastern (`western') sectors, consisting of eight regions each, originate at 10.5\,arcsec (193\,pc projected) from the nucleus and extend out to $300$\,arcsec ($\sim5.5$\,kpc projected). The conical diameter of the jet at the sector's base is $\sim3.7$\,arcsec ($\sim68$\,pc).}
	\label{fig:fluid-model-fig1}
\end{figure}

NGC\,5128/Centaurus\,A's interstellar medium (ISM) and its jets have been subject of numerous studies in X-rays with {\it Chandra} \citep{KRA00, KRA01, KRA02, KRA03, KRA08, KAR02, HAR03, HAR06, HAR07, KAT06, WOR08, CRO09, GOO10}. The extended X-ray emission from the ISM of the host galaxy NGC\,5128 has been well modelled as a thermal plasma with a $\beta$ model \citep{CAV76}, with $\beta\sim0.40$ between $\sim2$ and 11\,kpc projected distance from the nucleus \citep{KRA03}. {\it Chandra} imaging provides tight constraints on the jet width, and the data also allow us to place some limits on its bounding pressure.

The north-east-oriented jet that is currently active is traced out to $\sim5$\,kpc projected length in existing radio images (e.g.\,\citealp{HAR03, NEF15}); its X-ray counterpart (Fig.\,\ref{fig:fluid-model-fig1}) blends into the northern inner lobe at about 4.5\,kpc projected (e.g.\,\citealp{HAR06, HAR07}). No apparent disturbances occur in the galactic atmosphere surrounding the jet. On the south-western side, only a knotty structure up to $\sim2$\,kpc projected can be reliably associated with a jet with the current X-ray data (e.g.\,\citealp{HAR07}) and no diffuse X-ray emission from the counterjet has yet been seen. Where detected, the dominant X-ray radiation from both the diffuse emission and the knotty structures is unambiguously synchrotron (e.g.\,\citealp{HAR06, GOO10}). The jet is viewed at an angle to the line of sight of approximately 50$^\circ$ \citep{TIN98, HAR03}. Its opening angle, as measured from radio data, is $12^{\circ}$ on sub-pc and pc scales, and $15^{\circ}$ further out (e.g.\,\citealp{HOR06, GOO10, MUL14}). The brightness of the radio jets declines with distance from the nucleus more slowly than expected from an adiabatic jet, implying copious particle acceleration to compensate for the reduction in jet brightness from the expansion of the jet.

The jet exhibits apparent component speeds up to $\sim0.80c$ (intrinsic speed of $\sim0.63c$ for an inclination of $50^{\circ}$) at around $0.5$\,kpc projected, measured from radio data \citep{HAR03, GOO10}. \cite{SNI19} confirm these component speeds using X-ray data. With the measured apparent component speeds of $0.1-0.3c$ (intrinsic speed $\sim0.1-0.3c$) at subparsec scales \citep{TIN01, MUL14}, this points towards jet acceleration downstream (until it turns into a deceleration) or to sampling of different jet layers (discussed in general, in conjunction with a spine-sheath scenario, by \citealp{PIN18}). \cite{WOR08} show that the X-ray knots in the jet display a transverse trend in spectral index and so do not all lie at similar distances from the jet axis. Relying on sophisticated models for stellar mass loss into the jet, \cite{WYK15a} showed that Centaurus\,A's jet with power $\sim(1-2)\times10^{43}$\,erg\,s$^{-1}$ \citep{CRO09, WYK13, NEF15} can be slowed down to subrelativistic speeds with mass injection of $2.3\times 10^{-3}$\,M$_{\odot}$\,yr$^{-1}$.
\parskip 0.05cm

\cite{KRA03} speculated that the asymmetry of the inner lobes was induced by the differences in the environmental pressure of the host galaxy. If stellar material is entrained, the jet width is sensitive to initial and boundary conditions, so that relatively small changes in its surroundings can transform the jet into a lobe. This could account for the asymmetric morphology of Centaurus\,A on these scales (e.g.\,\citealp{KRA03}).

In the present paper, we rely on the pressure external to the Centaurus\,A jet, on the scale of the galactic atmosphere, derived from combined archival and new {\it Chandra} observations, and assume that the jet is approximately at pressure equilibrium with the bordering gas at any given point along its length. The fluid-like nature of the jet (owing to the transported magnetic fields) allows us to use an idealised, one-dimensional fluid model to calculate the runs of the jet velocity, energy distribution and mass-flow rate through the jet, to ascertain downstream parameters. Mass, energy and momentum conservation form the foundation of much of our analysis. The basic question is `Is there a self-consistent solution for the jet velocity, assuming local pressure equilibrium and mass input from stars alone (i.e.\,no external entrainment)? If so, what are the variations of velocity, density, mass flow, Mach number and so on along the jet?' The main novel features in the paper are the ability to use the known profiles of width and pressure in the model of the jet, and to model mass loss of realistic stellar populations tested against direct observational parameters.

The remainder of the paper is partitioned as follows. In Section\,\ref{sec:observations}, we document the X-ray observations, and describe the data reduction and analysis to obtain the principal physical parameters of the ISM encountered by the jet. Section\,\ref{sec:fluid-model} outlines the basic inputs for our analytic jet model. In particular, we provide appropriate conservation law expressions and outline our approach to extracting our proposed 1D jet flow model. Further, since the model depends on entrainment of local gas, it elucidates how we compute the physically-motivated stellar inputs such as the mass-return timescales for the NGC\,5128's stellar populations. In Section\,\ref{sec:solutions}, we present the solutions for a mass-loaded jet. We discuss the implications of our assumptions and of the findings in Section\,\ref{sec:discussion}, and conclude in Section\,\ref{sec:summary}. An appendix provides the details of the input parameters and some intermediate results.


\section{Data preparation and analysis} \label{sec:observations}

\subsection{\textit{Chandra} observations and data reduction} \label{sec:observations-reduction}

Previous analyses of the X-ray emission from the jet surroundings within about $6$\,kpc of the nucleus of NGC\,5128/Centaurus\,A revealed it to be dominated by thermal and synchrotron radiation, collectively peaking at energies below $1.0$\,keV (e.g.\,\citealp{KAR02, KRA03, KRA08, GOO10}). Since accurate spectral fitting of the soft X-ray band is required for our analysis, we opted for {\it Chandra} observations taken with the S3 chip of the Advanced CCD Imaging Spectrometer (ACIS) as it provides the greatest soft X-ray spectral sensitivity available with the instrument. 

The Centaurus\,A jet was initially observed with {\it Chandra} on 3 September 2002 with the target positioned on the S3 chip of ACIS in FAINT mode. Subsequent {\it Chandra} observations with identical telescope configuration were performed in 2003, 2009 and 2017, all of which centred on the nucleus. In a companion paper \citep{SNI19}, these observations are used to place constraints on morphological changes and proper motion of the X-ray bright knots in the jet. An overview of the observations used in the present work is given in Table\,\ref{table:obs}.

All data were reprocessed using {\sc ciao} v4.9, with the {\sc caldb} v4.7.6 calibration data base \citep{FRU06}. The {\sc ciao} task {\tt deflare} with default settings was run to dispose of background flares. The resulting cleaned exposure times, tabulated in Table\,\ref{table:obs}, total 264.5\,ks. 

As a next step, readout streaks in the images caused by the bright AGN core were removed using the task {\tt acisreadcorr}. The {\tt readout\_bkg} routine was employed to estimate the distribution of `out-of-time' events, those due to events that occur during frame transfer, for each observation. It is these cleaned exposures corrected for out-of-time events that we considered in the following analysis.
\begin{table}
	\caption{{\it Chandra} ACIS-S observations of Centaurus\,A used in this paper.}
	\label{table:obs}
	\begin{tightcenter}
	{\footnotesize 
		\begin{tabular}{ c c c }
		\hline
		ObsID & Date & $t_{\rm exp}^{a}$ \\
		 & & (ks)\\
		\hline
			02978 & 03-09-2002 & 44.6 \\
			03965 & 14-09-2003 & 48.9 \\
			10722 & 08-09-2009 & 49.4 \\
			19521 & 17-09-2017 & 14.8 \\
			20794 & 19-09-2017 & 106.8 \\
		\hline
		\multicolumn{2}{r}{Total exposure time} & 264.5 \\
		\hline
	\end{tabular}}
	\end{tightcenter}
	{${}^{a}$ Net exposure after background flare removal.}
\end{table}

To simulate a background event file for each observation, blank-sky exposures were taken from {\sc caldb}. Background rates were scaled to match observed rates in the 10--12\,keV energy band. The spectra utilised in the subsequent analysis were binned to have a minimum of 1 count per bin and were fitted over the energy range 0.6--2.0\,keV using the C-statistic (e.g.\,\citealp{CAS79, HUM09}), {\tt cstat} in {\sc xspec} v12.9.1k \citep{ARN96}. Abundances were scaled to the solar values of \citet{AND89}.

  
\subsection{Spectral analysis and deprojection of the ISM} \label{sec:deprojection}

In order to determine the physical properties of the ISM in the vicinity of Centaurus\,A's jet, we first carried out a spectral deprojection, assuming spherical symmetry of the ISM. The jet surroundings were divided into eastern and western sectors relative to the jet, with each sector having a base at $10.5$\,arcsec (193\,pc projected) from the nucleus (Fig.\,\ref{fig:fluid-model-fig1}). Those bases are upstream of where the first X-ray knots appear in the jet, but sufficiently far from the nucleus to avoid contamination by it, from the wings of the point-spread-function (PSF). Both pie slices were placed to trace the edge of the jet as closely as possible without encountering non-thermal contamination from the jet itself. Each sector was defined out to a galactocentric radius of $300$ arcsec ($\sim5.5$\,kpc projected) -- the X-ray jet itself is $\sim4.5$\,kpc long in projection, translating to a deprojected length at the inclination of $50^{\circ}$ of $\sim5.9$\,kpc -- and was adaptively divided into regions (annuli) with a minimum of 4000 counts over the 0.6--2.0\,keV band in each. In order to avoid contamination of the spectra, all the point sources coincident with the created sectors were masked to $\sim3$ times their FWHM, and the AGN core out to a radius of $10$ arcsec.

\begin{figure}
	\begin{tightcenter}
	\includegraphics[width=0.48\textwidth]{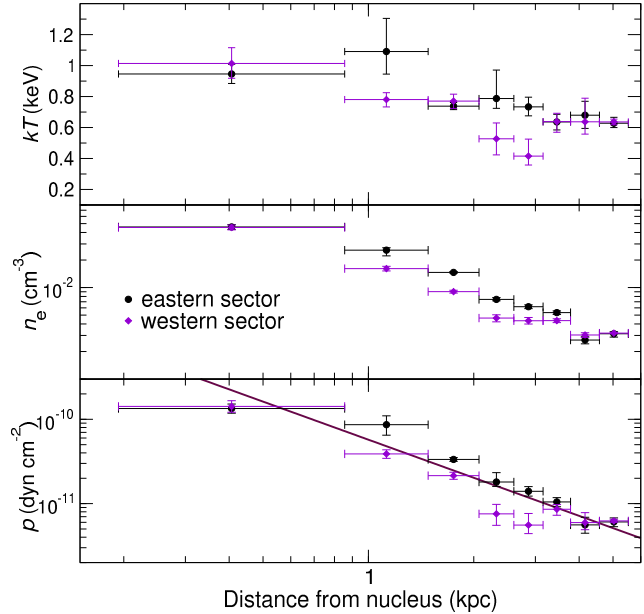}
	\end{tightcenter}
	\caption{Temperature, thermal electron number density and thermal pressure profiles obtained from the spectral deprojection, for the regions and sectors indicated in Fig.\,\ref{fig:fluid-model-fig1}. The circles (black) correspond to the eastern sector and the diamonds (purple) to the western sector. Vertical bars are $1\sigma$ uncertainties on the $\textsc{projct}$ values. The fit (solid line, red-brown) corresponds to the derived analytic expression (equation (\ref{eq:pressure})).}
	\label{fig:fluid-model-fig2}
\end{figure}
Each set of annular spectra was simultaneously fitted using the {\sc xspec} model $\textsc{projct}^*\textsc{phabs}^*\textsc{vapec}$. The {\sc vapec} thermal model \citep{SMI01} was selected because it allows the elemental abundances to be varied independently of one another. Two additional thermal components were included as a second model, to add small background corrections. The first set of these thermal components accounts for emission projected into the line of sight from regions outside the deprojection region, assuming that the gas is distributed as an isothermal $\beta$ model (\citealp{CAV76}; see \citealp{NUL10} for further details on this method of background correction). The $\beta$ parameter for the model (value $\sim 0.5$) was established by fitting the surface brightness profile over a $150$ to $300$ arcsec ($\sim2.8$ to $5.5$\,kpc) range (i.e.\,a cut, to avoid contamination from the dust lane on the lower end and the turnover region of the jet as seen in radio on the higher end). The second thermal component of the second model represents the thermal foreground emission from our Galaxy. Parameters of the \textsc{vapec} components were used to obtain the profiles in Fig.\,\ref{fig:fluid-model-fig2}.

The deprojection provides temperatures and abundances directly for spherical shells corresponding to the regions on the sky, while the electron densities are determined from the norms of the \textsc{vapec} thermal models, assuming that the density is uniform in the regions. Total pressures from the gas in and near Centaurus\,A are given by $n_{\rm tot}\,kT$, where the total adopted particle number density is $n_{\rm tot}\simeq1.93\,n_{\rm e}$. In measuring $n_{\rm e}$, abundances of oxygen, neon, magnesium and silicon were individually allowed to vary for each sector. All other elemental abundances were held fixed at 0.3\,Z$_{\odot}$ except for helium which was set to 1.0\,Z$_{\odot}$. For regions within the dust lane, hydrogen column densities were left free, while all regions outside the lane were frozen at the Galactic H\,{\small I} column density of $N_{\rm H}= 8.4\times10^{20}$\,cm$^{-2}$ (\citealp{DIC90}; consistent with \citealp{KAL05} who obtained $N_{\rm H}= 8.0\times10^{20}$\,cm$^{-2}$, within our uncertainties). 

The deprojected profiles are plotted in Fig.\,\ref{fig:fluid-model-fig2}, and the details of these results are provided in the appendix (Tables\,\ref{table:east} and \ref{table:west}). Temperatures in both profiles are elevated in the dust lane,\footnote{$N_{\rm H}$ was a free parameter in fits to establish the $1\sigma$ uncertainty on temperature.} less than 1.5\,kpc from the nucleus, but then converge to an average temperature of 0.65\,keV outside the lane. The temperature profiles separate between 2.5--3.5\,kpc, with the western sector decreasing in temperature by 30 per cent relative to the east. This temperature gradient could suggest the presence of a weak shock or some filamentary structure in the western sector; however, no feature was found through follow-up X-ray photometric and spectroscopic analyses. The electron densities in the two sectors follow similar declines with distance, although they can differ by up to 60 per cent. Deprojected pressures vary between the sectors by a factor of $\sim$2, yet good agreement is seen at large distances from the centre. We used the combined pressure results to derive an analytic expression for pressure $p$ in the range of $\sim0.2$ to 5.5\,kpc radial distance from the nucleus: 
\begin{equation} 
p(r) = (5.7\pm0.9) \times 10^{-11}\,(r/r_0)^{-1.5\pm0.2}\,{\rm\ dyn\,cm^{-2}}\,,
\label{eq:pressure}
\end{equation} 
where $r_0 = 1$\,kpc is the radial distance to normalize the gas density distribution. The analytic expression is represented by the solid line in Fig.\,\ref{fig:fluid-model-fig2}. The corresponding ISM mass density, for a constant ISM temperature of $0.65$\,keV (see also Table\,\ref{tab:keyvalues}), is compared to the mass density of the jet obtained from the one-dimensional fluid model in Fig.\,\ref{fig:fluid-model-fig3}; we turn our attention to the fluid model in Section\,\ref{sec:fluid-model}.

The thermal pressure in Fig.\,\ref{fig:fluid-model-fig2} can be considered to be the total pressure, as the contribution from magnetic fields and cosmic rays in the pc-kpc galactic atmosphere is negligible (e.g.\,\citealp{CRO09}). 

To relate to other observations, the ISM pressure 1.5--2\,kpc west from the centre (i.e.\,not coincident with our chosen sectors) as measured by \cite{CRO09} $\sim1.1\times10^{-11}$\,dyn\,cm$^{-2}$. It is roughly a factor of 2 lower than our deprojected value at the same distance from the nucleus. At a location corresponding to the outer regions ($\ga3.5$\,kpc) of our eastern sector, the pressure from \cite{KRA08} is approximately $3.0\times10^{-12}$\,dyn\,cm$^{-2}$, again a factor of about $2$ lower than found in this work.


\section{Jet fluid model}  \label{sec:fluid-model}

Our objective is to make a quantitative model of flow through the jet that matches observed properties of the jet and its environment, and examine under what conditions it is possible to decelerate a relativistic core jet to a subrelativistic flow. Here, we outline a physical model for this purpose. 

The Centaurus\,A jet is fairly well collimated and, apart from the region close to the brightest X-ray knots, AX1A and AX1C (see \citealp{GOO10, SNI19}) at about 350\,pc (deprojected), its width varies smoothly with distance from the nucleus; the modest opening angle of $\sim15^{\circ}$ (Section\,\ref{sec:intro}) motivates an assumption of paraxial plasma flow. This indicates that the jet is pressure-confined over most of its length (Section\,\ref{sec:equations}) and that the jet flow does not vary rapidly with time. These properties suggest the use of a steady, one-dimensional flow model. Although such a model is clearly approximate, it can provide estimates of flow properties and some insight into the behaviour of the jet.

Knotty substructures within the jet reveal local departures from the steady, one-dimensional flow.  Following previous work \citep{HAR03, WYK15a}, we presume that the bulk of these knots are sites where the jet interacts with stellar winds, which leads to dissipation and turbulence. The turbulence likely adds to the effective pressure of the jet fluid, but we assume that any turbulent pressure can be lumped together with the `thermal' pressure of the jet plasma. The one-dimensional model requires the turbulence and dissipation to be locally uniform when averaged over regions approaching the width of the jet. Again, this approximation will be poorest in the vicinity of knots AX1A and AX1C. However, provided that mass, momentum and energy are conserved, the model can be used to bridge across regions where our detailed assumptions may not be accurate.
\begin{figure}
	\begin{tightcenter}
	\includegraphics[width=0.53\textwidth]{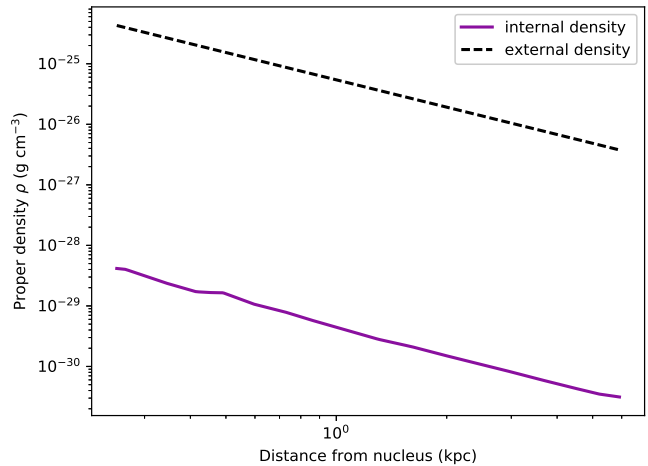}
	\end{tightcenter}
	\caption{Jet internal mass density (solid purple line) and ISM mass density (dashed black line) as a function of deprojected distance from the nucleus. Here, the ISM density is determined from the pressure given by equation (\ref{eq:pressure}), assuming the ideal gas law and a constant temperature of $kT = 0.65$\,keV. The glitches upstream in the jet, seen in this and the following figures, coincide with the `flaring region' at around $350$\,pc from the nucleus, associated with the base (A1) knots (see e.g.\,\citealp{SNI19}); the model is unreliable around this point.}
	\label{fig:fluid-model-fig3}
\end{figure} 

Almost certainly, the flow speed varies to some degree with distance from the jet axis, violating our assumption of a one-dimensional flow. We remark that, between projected distances of $\sim3$ and $4.5$\,kpc from the nucleus, the X-ray jet tapers from the full width of the radio jet to a sharp tip, before disappearing. If the production of X-ray synchrotron emission depends primarily on the flow speed, this abrupt disappearance of the X-ray jet requires either a steep velocity gradient in this region, or that the X-ray synchrotron emissivity is very sensitive to the flow speed. The weak dependence of the velocity on distance from the AGN nucleus found in our models (see Section\,\ref{sec:solutions}) would require the latter. The taper at the end of the X-ray jet implies there is some gradient in flow speed from the spine of the jet to its sheath. However, if particle acceleration is very sensitive to the flow speed, the required speed difference will be modest. Thus, while the taper does imply some transverse velocity gradient in the jet, the one-dimensional flow model should provide good, representative estimates of the flow properties.

We disregard plasma instabilities as these occur only on scales of order the gyroradius. But, we need to be mindful of fluid instabilities, despite them not being an actual input in the model; they are relevant in the context of external entrainment and in the context of potential disruption of the jet. \cite{BLA74} pointed out that small-scale Kelvin-Helmholtz (KH) instabilities of the type endemic to jet flows could have a short growth time, and grow so rapidly that the overall mean jet flow could be regarded as steady. We have evaluated the length scales and growth rates of KH modes for our model by solving the dispersion relation from \cite{BIR84, BIR91}. The internal medium is taken to have a density $3 \times 10^{-4}$ of the density of the external medium (Fig.\,\ref{fig:fluid-model-fig3}). The internal and external sound speeds, for pressure balance, are then about $10^5$ and $2 \times 10^3$\,km\,s$^{-1}$. Such a light and fast flow is relatively stable against the ordinary $n = 0$ (`pinching') type modes, but quite unstable to the ordinary $n=1$ (`helical') and higher-$n$ (`fluting') modes and the reflection modes of all $n$. The flow supports a large number of unstable ordinary or reflection modes of short wavelengths, less than about $10$ jet radii, and growth lengths of a few jet radii, with the growth length decreasing for higher-order modes of shorter wavelength. The implication is that the effect of the KH instability is to cause a jet flow initially bounded by a sharp velocity and density discontinuity to develop smoother velocity and density profiles, rather than to be disrupted. This is consistent with the lack of observational evidence for large-scale KH instabilities in Centaurus\,A. The region affected has to be quite thin. The agreement between the X-ray and radio sizes (see Section\,\ref{sec:methodology}) also argues that any transition layer between the jet and ISM must be narrow; otherwise, the radio jet should be wider than the X-ray jet.



\subsection{Fluid equations} \label{sec:equations}

If energy is conserved, the kinetic energy of any entrained matter, as measured in the jet rest frame, is dissipated in the jet fluid. We assume that entrainment of stellar mass loss is the primary cause of dissipation within the jet, requiring mass entrainment to be incorporated into the model flow equations (e.g.\,\citealp{KOM94}). Assuming that the particle number in the jet is conserved, the flow of matter through it can be tracked in terms of the proper (i.e.\,measured in the fluid rest frame) density of rest mass $\rho_{\rm j}$. The (smoothed) rate per unit volume at which stars inject rest mass into the jet is a relativistic scalar density, which we denote by $\alpha$. The conservation of matter is then expressed by the continuity equation
\begin{equation}
  \frac{\partial}{\partial x^\mu}\, \rho_{\rm j}\, U^\mu = \alpha\,,
\end{equation}
where $x^\mu = (ct, \mathbf{r})$ refers to the 4-position and $U^\mu = \Gamma_{\rm j} (c, \mathbf{v})$ is the 4-velocity of the jet fluid. For steady flow, this reduces to 
\begin{equation} 
  \nabla \cdot \rho_{\rm j}\, \Gamma_{\rm j}\, \mathbf{v} = \alpha\,,
\label{eq:scont}
\end{equation}
where $\nabla \cdot$ is the three-space divergence.
If the one-dimensional coordinate on which the jet properties depend is $s$, we consider a volume $\mathcal{V}$ of the jet that is bounded on its sides by the edges of the jet and on its ends by surfaces of constant $s$, say $s = s_1$ and $s_2$ at the inner and outer ends, respectively. Integrating equation (\ref{eq:scont}) throughout $\mathcal{V}$ and using the divergence theorem gives
\begin{equation} 
  \int_{\partial \mathcal{V}} \rho_{\rm j}\, \Gamma_{\rm j}\, \mathbf{v} \cdot {\rm d}\mathbf{A}
  = \int_\mathcal{V} \alpha \, {\rm d}\mathcal{V}\,,
\label{eq:mass}
\end{equation}
where $\partial \mathcal{V}$ is the boundary of $\mathcal{V}$. Since the flow velocity $\mathbf{v}$ is parallel to the sides of the jet, they do not contribute to the surface integral. If $A(s)$ is the area of the surface within the jet on which the coordinate has the value $s$, the flux of rest mass through this surface is
\begin{equation}
  \dot{M}(s) = \rho_{\rm j}(s)\, \Gamma_{\rm j}(s)\,v_{\rm j}(s)\, A(s)\,,
\label{eq:mdot}
\end{equation}
where $v_{\rm j}(s)$ denotes the flow speed through the surface (the values $\rho_{\rm j}(s)$, $\Gamma_{\rm j}(s)$ and $v_{\rm j}(s)$ are well defined, since the flow is one-dimensional). Thus, equation (\ref{eq:mass}) gives
\begin{equation} 
\dot{M}_2 - \dot{M}_1 = \int_\mathcal{V} \alpha \, {\rm d}\mathcal{V}\,,
\label{eq:continuity}
\end{equation}
where $\dot{M}_1 = \dot{M}(s_1)$ and $\dot{M}_2 = \dot{M}(s_2)$.

We assume that the fluid is perfect (has isotropic stresses in its local rest frame), so that the stress-energy tensor has the form \citep{LAN59}
\begin{equation} 
  T^{\mu\nu} = w\, U^\mu\, U^\nu / c^2 + p\, g^{\mu\nu},
\label{eq:emom}
\end{equation}
where the Minkowski metric is $g^{\mu\nu} = \mathop{\rm diag}(-1, 1,
1, 1)$, $p$ is the proper pressure and $w = e + p$ is the proper enthalpy density, with $e$ the proper energy density (including rest mass). In the transrelativistic range considered here, it is appropriate to
partition the enthalpy as
\begin{equation} 
  w = \rho_{\rm j}\, c^2 + h\,,
\label{eq:enthalpy}
\end{equation}
so that $h$ asymptotes to the more familiar, non-relativistic enthalpy in the low-energy limit. The energy-momentum injected into the jet fluid per unit of stellar mass loss can be expressed in the form $\epsilon V^\mu / c$, where $\epsilon$ is the proper specific energy (energy per unit mass) in the stellar winds, and $V^\mu$ is a time-like unit 4-vector, in which case the equation for conservation of energy-momentum takes the form
\begin{equation} 
  \frac{\partial}{\partial x^\mu} T^{\mu \nu} = \alpha\, \epsilon\, V^\nu
  / c^2\,.
\label{eq:SE}
\end{equation}
In the rest frame of the host galaxy, the frame in which the flow is steady, the net momentum introduced by stellar mass loss is small and has little impact, so we take it to be exactly zero.\footnote{On the other hand, stellar mass entrained into the jet is substantial; hence, the `0'-momentum component, $mc$, is not neglected.} Some thermal energy will also be introduced to the jet with the stellar winds, but this is negligible compared to the thermal energy liberated by mixing stellar wind into the fast-moving jet. Therefore, in the galaxy frame, we assume $V^\mu = (c, 0, 0, 0)$ and $\epsilon = c^2$. With these assumptions, for steady flow, equations (\ref{eq:emom}) and (\ref{eq:SE}) give 
\begin{equation} 
  \nabla \cdot \left[\,\frac{w}{c^2}\, \Gamma_{\rm j}^2\, \mathbf{v} \begin{pmatrix} c \\ \mathbf{v}
  \end{pmatrix} \right] + \begin{pmatrix} 0 \\ \nabla p \end{pmatrix} 
  = \alpha \begin{pmatrix} c \\ \mathbf{0} \end{pmatrix}\,.
\label{eq:energmom}
\end{equation}
We neglect anisotropic magnetic stresses.

The upper component of equation (\ref{eq:energmom}) expresses conservation of energy for the steady flow. Integrating it throughout the volume $\mathcal{V}$ described above gives
\begin{equation}
  \left[ A\, w\, \Gamma_{\rm j}^2 v_{\rm j} \right]_1^2
  = \int_\mathcal{V} \alpha\, c^2 \, {\rm d}\mathcal{V} = \left[ \dot{M} c^2 \right]_1^2,
 \label{eq:eq11}
\end{equation}
where equation (\ref{eq:continuity}) has been used on the right-hand side. Moving the terms from the right-hand side to the left and using equations (\ref{eq:enthalpy}) and (\ref{eq:mdot}), we find that the jet power
\begin{equation}
  P_{\rm j} = A\, w\, \Gamma_{\rm j}^2\, v_{\rm j} - \dot{M} c^2
  = (\Gamma_{\rm j} - 1) \dot{M} c^2 + A\, h\, \Gamma_{\rm j}^2\, v_{\rm j}
 \label{eq:power}
\end{equation}
remains constant in the jet. The first term on the right is the kinetic power and the second term gives the power in internal energy carried by the jet (`thermal power'). Note that the conserved jet power does not include rest-mass energy, since that varies as mass is entrained by the jet.

The lower (3-space) components of equation (\ref{eq:energmom}) contain the usual momentum equation
\begin{equation} 
  \nabla \cdot \frac{w}{c^2}\, \Gamma_{\rm j}^2\,\mathbf{v} \mathbf{v}  + \nabla p = 0\,,
 \label{eq:mom}
\end{equation}
where $\mathbf{v} \mathbf{v}$ is a dyadic (or tensor) product. The equation requires the pressure to be continuous across the jet boundary. In equation (\ref{eq:mom}), the only contribution of the first term that need not be parallel to $\mathbf{v}$ is proportional to $\mathbf{v} \cdot \nabla \mathbf{v}$. Thus, the pressure gradient will be parallel to $\mathbf{v}$, except where the streamlines have significant curvature. Once more, the abrupt expansion of the jet in the vicinity of knots AX1A and AX1C means that at least some streamlines are  strongly curved there, so that the pressure may vary signficantly across the streamlines. Not surprisingly, our assumption of one-dimensional flow is likely to be poorest in this region. We also point out that our assumption of one-dimensional flow requires the jet to be irrotational: with $\nabla s$ the flow direction, the velocity is expressible as $v_{\rm j} = f(s) \nabla s$, from which it follows that $\nabla \times \mathbf{v} = 0$. In practice, the large expansion during outflow from the core will reduce any initial circulation around the jet axis. Some circulation is expected due to the speed difference between the spine and sheath, but, as discussed in Section\,\ref{sec:fluid-model}, that is probably not very large.

Because the jet speed is much greater than the free-fall speed across the range of distances of interest, the direct effect of gravity on the jet velocity and pressure will be negligible. The only significant external force on the jet is the net pressure force from the surrounding regions, which {\it is} included in the model.

Taking the dot product with a constant vector $\mathbf{b}$, integrating throughout the volume $\mathcal{V}$, and using the divergence theorem, equation (\ref{eq:mom}) yields
\begin{align} 
  0 &= \int_{\partial \mathcal{V}} \frac{w}{c^2}\, \Gamma_{\rm j}^2 (\mathbf{v} \cdot \mathbf{b})
  \mathbf{v} \cdot {\rm d}\mathbf{A} + \int_\mathcal{V} \mathbf{b} \cdot \nabla p \, {\rm d}\mathcal{V} \nonumber \\ 
  &= \left[ \left. A \frac{w}{c^2}\, \Gamma_{\rm j}^2\, v_{\rm j}\, \mathbf{v} \right|_1^2
  + \int_\mathcal{V} \nabla p \, {\rm d}\mathcal{V} \right] \cdot \mathbf{b}\,,
\label{eq:buoy}
\end{align}
where, again, there is no contribution to the surface integral from the sides of the jet, because $\mathbf{v}$ is parallel to the surface. Since $\mathbf{b}$ is arbitrary, the vector in brackets must be zero. Its first term is the increase in the jet momentum flux between the surfaces at $s = s_1$ and $s = s_2$, while the second term is minus the net pressure force on the jet in the volume $\mathcal{V}$.

Henceforth, we assume that the coordinate $s$ can be taken to be the radial distance $r$ from the nucleus to the point of interest. Since the opening angle of the jet is modest, the upward component of the momentum flux through the surface $A$ normal to the jet axis is close to
\begin{equation} 
  \Pi = A\,w\, \Gamma_{\rm j}^2\, \beta_{\rm j}^2 = (P_{\rm j} / c + \dot{M} c)\, \beta_{\rm j}\,,
\label{eq:momflux}
\end{equation}
where $\beta_{\rm j} = v_{\rm j}/c$, while equations (\ref{eq:enthalpy}), (\ref{eq:power}) and (\ref{eq:mdot}) have been used in turn to eliminate $w$, $h$ and $\rho_{\rm j}$ to obtain the expression on the right. Most of the terms in equation (\ref{eq:momflux}) are functions of $r$. The approximation in equation (\ref{eq:momflux}) amounts to replacing the average value of $\cos\psi$ over a level surface of $s$ by unity. For a conical jet with full opening angle $\psi$, the actual average value is given by $\cos^2 (\psi/4) \simeq0.996$ for an opening angle of $15^{\circ}$. Similarly, the second term of the momentum equation can be approximated by the radial pressure gradient, giving the increase in momentum flux due to the external force, i.e.\,the net pressure force, on the jet as
\begin{equation} 
  \Pi_2 - \Pi_1 = - \int_\mathcal{V} \frac{{\rm d} p_{\rm ISM}}{{\rm d} r} A(r) \, {\rm d} r\,.
\label{eq:pforce}
\end{equation}
Note that this result is exact if the flow is spherical. Below, thrust is used to mean the momentum flux of the jet.

Before discussing further details, we reiterate that our flow solution only relies on assuming that particle numbers and energy are conserved. If the input parameters are reasonable, and key properties, such as the jet power, have remained nearly constant over the $\sim5\,$kpc$/0.5c\sim 3\times 10^4$\,yr required for the jet to flow through the region of the solution, it should provide representative results. The model may not be accurate in parts of the jet where some of our assumptions are not well satisfied, but this will not cause it to fail in other regions where the assumptions are better met.



\subsection{Model implementation} \label{sec:methodology}

Using the results of the previous subsection, we model the jet by solving equation (\ref{eq:power}) for the jet speed. There are several parameters that must be evaluated in order to do this.

We adopt the jet power of $P_{\rm j}\sim1\times10^{43}$\,erg\,s$^{-1}$ from \cite{CRO09}. This power estimate is based on shock dynamics and should be mostly independent of the jet composition. It is uncertain by a factor $\sim2$.\footnote{A factor 2 seems reasonable to account for the uncertainties in the external pressure, the geometry of the inner lobes (including projection), the assumption of a constant speed of lobe expansion, and the inner lobe age (also including projection), which are not estimated by \cite{CRO09}.} We argue that the power remains approximately constant along the jet length (see equation (\ref{eq:power})). This is reasonable given the small power radiated and the apparent absence of disturbances surrounding the jet (Section\,\ref{sec:intro}).

The area of the jet at a given distance from the nucleus was deduced from our {\it Chandra} images of the radio galaxy (except that we use Very Large Array, VLA, data where the X-ray jet tapers off at its most downstream region). We measured the transverse diameter of the X-ray or radio jet from the images at 15 locations along its length, where the diameter is taken to be the broadest extent of any detectable X-ray or radio emission. Our measurement accuracy corresponds to approximately $\pm 0.5$ {\it Chandra} pixels, or $0.25$ arcsec, for the X-ray data, and is similar for the radio data. Here, the implicit assumption is that the physical jet flow is not significantly larger than the region of observable radio or X-ray emission; it cannot be smaller and we have no reason to expect that it is larger. The fact that the X-ray and radio diameters are in good agreement suggests that there is no bias in using the X-ray data. The opening angles corresponding to these measurements are given in Table\,\ref{tab:locations} (this sampling rate is adequate, as adding more points only increases the apparent noise in the solutions). Intermediate values are found by linear interpolation. The flow equations are solved at any nominated location, as required. 
\begin{figure}
	\begin{tightcenter}
	\includegraphics[width=0.53\textwidth]{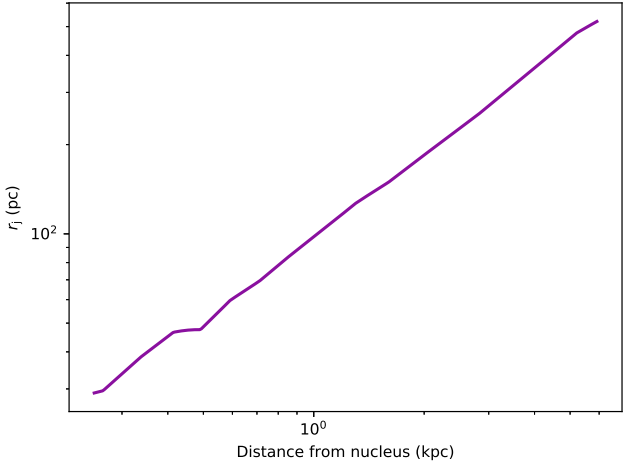}
	\end{tightcenter}
	\caption{Jet radius as measured from {\it Chandra} images (see the text) as a function of deprojected distance from the nucleus for a jet inclination of $50^{\circ}$. The near-linearity suggests a constant opening angle, with a value of $\psi\sim 5.68/0.49 \sim 11.6^{\circ}$.}
	\label{fig:fluid-model-fig4}
\end{figure} 

To determine the jet transverse area $A(r)$, we assume that the projected radius of the jet at the projected distance $r \sin\theta$ from the nucleus is equal to the actual jet radius $r_{\rm j}$ (X-ray to zero intensity) at $r$ (Fig.\,\ref{fig:fluid-model-fig4}). Here, $\theta$ is the inclination of the jet to our line of sight. This approximation is accurate, provided that the projected opening angle of the jet $\psi$ is modest. We also use the planar approximation, $A(r) = \pi r_{\rm j} (r)^2$, for the jet area. If the jet flow is radial, the total error due to these approximations would be $\la 1$ per cent, which is negligible compared to other sources of error. 

The pressure in the jet is taken to match the pressure profile given by equation (\ref{eq:pressure}). Assuming a fixed ratio of specific heats $\gamma$, this determines the enthalpy density as
\begin{equation}
  h = \frac{\gamma}{\gamma - 1}\, p\,.
\end{equation}
The value used for the ratio of specific heats, $\gamma = 13/9$, is appropriate for hydrogen plasma in the temperature range $0.5{\rm\ MeV} \la kT \la 1 {\rm\ GeV}$, when the electrons are relativistic, but the protons are not\footnote{The net ratio of specific heats is $(n_{\rm e}\,c_{\rm p,e} + n_{\rm p}\,c_{\rm p,p}) / (n_{\rm e}\, c_{\rm \mathcal{V},e} + n_{\rm p}\, c_{\rm \mathcal{V},p})$, where $n_{\rm e}$ and $n_{\rm p}$ are the number densities of electrons and protons, respectively, $c_{\rm p,e}$ and $c_{\rm p,p}$ are the respective specific heats per particle at constant pressure, and $c_{\rm \mathcal{V},e}$ and $c_{\rm \mathcal{V},p}$ are the specific heats at constant volume. For hydrogen plasma, $N_{\rm e} = N_{\rm p}$, so the calculation reduces to $(4 + 5/2) / (3 + 3/2)$. $13/9$ is sensible in our case, where it can be assumed that the electrons and protons are in sufficiently close thermal contact that they act as a single coupled thermal fluid.} (relativistic electrons, subrelativistic protons, where by number $N_{\rm e}\sim N_{\rm p}$ and positrons are already diluted to insignificance). As shown in Fig.\,\ref{fig:fluid-model-fig9}, this is the effective temperature range we find for the jet plasma.

Our assumptions so far determine everything but the jet speed and $\dot{M}(r)$ in equation (\ref{eq:power}). To complete the solution, we need a value for one or the other of these. At the innermost solution point, in the absence of other constraints on the flux of rest mass, we specify the
jet speed. Equation (\ref{eq:power}) then decides the value of $\dot{M}(r)$ there. Knowing $\dot{M}(r)$ at the first solution point and $\alpha(r)$, equation (\ref{eq:continuity}) can be used to determine $\dot M(r)$ at any other point. Knowing $\dot{M}(r)$, equation (\ref{eq:power}) can now be solved for the flow speed. The rate of stellar mass injection $\alpha(r)$ is discussed in the following sections.

Having determined values for $\dot{M}(r)$ and $v_{\rm j}(r)$, equation (\ref{eq:mdot}) can be solved for the proper density of rest mass in the jet plasma $\rho_{\rm j}(r)$. Combining that with the jet pressure in the ideal gas law, we can establish an effective temperature $kT$ for the jet plasma. For this purpose, we assume a mean mass per particle of $0.59\,m_{\rm H}$, typical of fully-ionized gas with cosmic abundances. Although the particle distribution is unlikely to be thermal, this value of $kT$ should be representative of the particle energies measured in the rest frame of the jet plasma.  Other jet properties, such as the powers in internal and kinetic energy, and the sound speed of the jet plasma, can be ascertained from these.

We consider `initial' speed (speed at the start of the modelled region, 193\,pc projected distance from the nucleus) of $v_{\rm j} = 2c/3 = 0.667c$ (and so a Lorentz factor $\Gamma_{\rm j}=1.34$) and inclination of the approaching jet $\theta = 50^{\circ}$. After an initial exploration, we consider the effects of varying the initial jet speed, $P_{\rm j}$ and $\theta$, as these are less well constrained than other parameters. As required by the one-dimensional flow model, the velocity and density are taken to be constant across the jet. 

The steady one-dimensional flow solution is fully determined by the procedure above, so that the momentum equation (\ref{eq:pforce}) can be used as a consistency check on the solution. In general, the momentum flux, equation (\ref{eq:momflux}), will depend on the galactocentric radius $r$ in a flow solution. It is possible that the jet loses an appreciable amount of momentum due to effective viscous stresses at the jet boundary, but it is very unlikely that the jet momentum flux could increase by more than the amount due to the net pressure force, given by equation (\ref{eq:pforce}). As discussed below, this proves to be a significant constraint.

We remark in addition that the jagged features visible in the plots of the flow solutions, due to their limited dynamic ranges, most evident in the temperature and Mach number (Figs\,\ref{fig:fluid-model-fig9} and \ref{fig:fluid-model-fig10}), reflect our discrete measurements of the jet width (Table\,\ref{tab:locations}). At small radii, these are most affected by the flare associated with knots AX1A and AX1C at $\sim350$\,pc deprojected distance; almost certainly, these knots play some role in producing the temperature peak seen in this distance range in Fig.\,\ref{fig:fluid-model-fig9}. More generally, the measurements are also affected by knots near the jet margin and regions where the edge of the jet is less well defined.

We wrote {\sc python} codes to obtain the simulated jet velocity profile, power distribution and mass-flow rate.\footnote{The codes used for this paper are available from the authors upon reasonable request.} These simulations are not computationally expensive, and we used a personal platform running Qubes 4.0.\footnote{\tt https://www.qubes-os.org/} More demanding, additional simulations to assess stellar mass-loss rates (Section\,\ref{sec:mass-return}) were conducted on the University of Hertfordshire cluster.\footnote{\tt https://uhhpc.herts.ac.uk/}

\subsubsection{ISM parameters} \label{sec:ISM}

The ISM gas temperature is set to the mean value of $0.65$\,keV (Section\,\ref{sec:deprojection}) and the pressure profile of the jet is assumed to be as specified in equation (\ref{eq:pressure}). We note that this temperature is used solely to determine the density of the ISM from the pressure in Fig.\,\ref{fig:fluid-model-fig2}. Distance along the jet is related to the observed (projected) distance, assuming a jet inclination of $50^\circ$.

We need to determine the mass injection rate $\alpha$. In the following sections, we proceed by estimating the total density of gravitating matter and, from that, the stellar density. A model for
the stellar population is then used to determine the rate at which the stars shed mass.

\subsubsection{Stellar mass density} \label{sec:fstar}

The gravitating mass distribution is approximated as singular isothermal sphere, with Keplerian velocity $v_{\rm K}$, requiring the gravitating mass density given by
\begin{equation}
\rho_{\rm grav} = \frac{v_{\rm K}^2}{4 \pi G r^2}\,.
\end{equation}
This approximation is satisfactory, although it becomes poorer near the innermost solution point. We use $v_{\rm K} = 250$\,km\,s$^{-1}$ from \cite{GRA79}, and other authors (e.g.\,\citealp{HUI95}) obtain similar results.

We take $75$ per cent of the stellar population of NGC\,5128 to be 12\,Gyr-old stars and the remaining $25$ per cent to be 3\,Gyr old \citep{REJ11, WYK15a}, or in the terminology of the model used below, $f_1 = 0.75$ and $f_2 = 0.25$.\footnote{The percentages of $\sim75$ per cent of old ($\sim12$\,Gyr) and $\sim25$ per cent of younger ($\sim3$\,Gyr) stars are based on simulated colour-magnitude diagrams and refer to the percentage of stars. While not strictly equal to mass fractions, given the relatively narrow range of masses of surviving RGB stars that were probed, this is close to mass-based grouping (see \citealp{REJ11}).}

The stellar $M/L$, in $V$-band, in the BaSTI population synthesis models \citep{PIE04, PIE06}\footnote{\tt http://albione.oa-teramo.inaf.it/} for the 12\,Gyr population (alpha-enhanced, metallicity $Z$ = 0.004 and mass-loss efficiency parameter\footnote{$\eta$ is defined by \cite{REI75}, and \cite{MCD15} provide its value.} $\eta = 0.4$) is $2.73$, and for the 3\,Gyr population (solar-scaled, $Z$ = 0.008, $\eta = 0.4$) the $M/L_V$ is $1.16$. The younger stars will reduce the composite value:
\begin{equation}
M/L = \frac{f_1 \rho_\ast + f_2 \rho_\ast}{f_1 \rho_\ast / \mu_1 + f_2 \rho_\ast / \mu_2}
 = \frac{f_1 + f_2}{f_1 / \mu_1 + f_2 / \mu_2}\,,
\end{equation}
where $f_1 + f_2 =1$, $\rho_*$ is the stellar mass density, and $\mu_1$ and $\mu_2$ are the $M/L$ of the older and younger populations respectively. From the above stellar population $M/L_V$ values, we obtain a composite modelled $M/L_V$ of about $2.04$. 

The measured (i.e.\,including dark matter) $M/L_V$ for NGC\,5128/Centaurus\,A can be retrieved from \cite{HUI95}. An appropriate value for our modelled region ($4.5$\,kpc projected) follows from their figure $21b$ showing estimated $M/L_B$, and converts to $M/L_V \sim3.8$ for our adopted distance to Centaurus\,A. If the visible light comes from this population of stars, the discrepancy between the stellar modelled (BaSTI) $M/L_V$ ratio and the observed value must be due to the presence of dark matter. The ratio of the predicted $M/L_V$ to the observed value then gives the fraction of the gravitating mass in stars, $f_* \simeq 2.04 / 3.8 \simeq 0.54$. The mean stellar density is then
\begin{equation}
\rho_* = f_*\, \rho_{\rm grav}\,. 
\end{equation}

\subsubsection{Mass-return timescale} \label{sec:mass-return}

Averaged over the population of stars, the mean rate per unit volume at which the stars shed mass can always be expressed in the form
\begin{equation}
\alpha = \frac{\rho_*}{\tau}\,,
\label{eq:alpha}
\end{equation}
where $\tau$ is called the mass-return timescale, 
\begin{equation}
\tau = \frac{M_*}{{\rm d}M_* / {\rm d}t}\,,
\end{equation}
with $M_*$ the mass of a representative population of stars at time $t$. Here, we discuss the appropriate value of $\tau$ for the stars in NGC\,5128.

For a population of stars born in a single event, the total stellar mass-loss rate is
\begin{equation}
\frac{{\rm d}M_{\rm tot}}{{\rm d}t} = \Big[\,\frac{{\rm d}N}{{\rm d}t}\,\Big]_{M_{\rm init}} (M_{\rm init} - M_{\rm fin})\,, 
\label{eq:masslossrate}
\end{equation}
where $M_{\rm tot}$ is the total mass of the population, $\frac{{\rm d}N}{{\rm d}t}|_{M_{\rm init}}$ is the stellar death rate, evaluated at $M_{\rm init}$, and $M_{\rm init}$ and $M_{\rm fin}$ refer to the considered initial and final stellar masses. The term ${\rm d}N/{\rm d}M_{\rm init}$ comes from the initial mass function (IMF) and the term ${\rm d}M_{\rm init}/{\rm d}t$ from stellar evolution models of the change in stellar lifetime with mass.

Writing equation (\ref{eq:alpha}) for individual populations, we have
\begin{equation}
\alpha = \frac{f_1 \rho_\ast}{\tau_1} + \frac{f_2 \rho_\ast}{\tau_2} = \frac{(f_1 + f_2) \rho_*}{\tau}\,,
\end{equation}
where $\rho_*$ is again the stellar mass density, $\tau_1$ and $\tau_2$ are the mass-return timescales of the 12\,Gyr and the 3\,Gyr populations, respectively, and $\tau$ is the composite result.

\cite{FAB76} suggest an overall mass-loss rate in large ellipticals of $1.5 \times 10^{-11}$\,M$_{\odot}$\,yr$^{-1}$\,L$_{\odot}^{-1}$. Since we have a reasonable understanding of the stellar content of NGC\,5128, we can compare the mass-loss rates and work out the mass-return timescales in a more detailed fashion. 

We use the stellar evolution code described by \cite{HUR00} and stellar wind codes by \cite{CRA11} (see \citealp{WYK15a} for the details on the code handling), with a modification to compute the mass-return timescale. In brief, the Single-Star Evolution (SSE) routine by \cite{HUR00} is based on a number of interpolation formulae as a function of the initial mass, stellar age and metallicity, and provides predictions for $\dot M$ for phases with high mass-loss rates. The BOREAS routine by \cite{CRA11} is added to fill in for the missing mass-loss rates; this routine computes $\dot M$ for cool main-sequence stars and evolved stars (red giant branch, RGB, not asymptotic giant branch, AGB).

Again, we consider 75 per cent of 12\,Gyr ($Z = 0.004$) and 25 per cent of 3\,Gyr-old ($Z=0.008$) stars. We adopt the IMF as used by \cite{WYK15a}: $x = 1.3$ between $0.08$ and $0.5$\,M$_{\odot}$ and $x = 2.35$ for $0.5$\,M$_{\odot}$ and higher masses ($x$ in the sense $M_{\rm init}^{-x}$). $10^8$ stars are simulated to avoid small-number effects associated with brief stages of substantial mass-loss rate at the tip of the AGB, in $20$ runs. We model the $12$ and $3$\,Gyr populations separately, and compare with direct observations. The retrieved mass-return timescale $\tau$ is independent of the jet opening angle; it only depends on the stellar population properties and the assumptions about stellar evolution.

For the 12\,Gyr ($Z = 0.004$) population, we obtain a mass-loss rate per unit luminosity of $(7.17 \pm 0.14) \times 10^{-12}$\,M$_{\odot}$\,yr$^{-1}$\,L$_{\odot}^{-1}$ and a mass-loss rate per unit mass of $(3.00 \pm 0.07) \times 10^{-12}$\,M$_{\odot}$\,yr$^{-1}$\,M$_{\odot}^{-1}$. The latter gives a mass-return timescale of $\tau_1\sim3.33 \times 10^{11}$\,yr. 

While the mass-loss rate is somewhat below the value for old populations in large ellipticals given by \cite{FAB76}, we have direct observational evidence for a similar mass-return timescale in the nearby stellar cluster 47\,Tucanae,\footnote{Stellar mass-loss rate is independent of the environment.} which harbours a single population of $11.95$\,Gyr-old stars \citep{MCD15} with metallicity of $Z = 0.003$ \citep{ROE14}. Its total mass is $M_{\rm tot} = 1.1 \times 10^6$\,M$_{\odot}$ \citep{LAN10} and the stellar death rate amounts to 1 per 80\,kyr \citep{MCD11}. The initial mass of a star $M_{\rm init}$ has been estimated as $0.89$\,M$_{\odot}$ \citep{MCD15, FU18}, while the final mass is $M_{\rm fin} = 0.53$\,M$_{\odot}$ \citep{KAL09}. Following equation (\ref{eq:masslossrate}), the resulting rate amounts to $(1/8\times 10^4)\times (0.36/1.1\times 10^6) \sim 4.09 \times 10^{-12}$\,yr$^{-1}$, i.e.\,a mass-return timescale $\tau_1\sim2.44 \times 10^{11}$\,yr, which is fairly close to the modelled value above. 

Separately modelling the $3$\,Gyr ($Z=0.008$) population gives a mass-loss rate per unit luminosity of $(1.57 \pm 0.09) \times 10^{-11}$\,M$_{\odot}$\,yr$^{-1}$\,L$_{\odot}^{-1}$ and a mass-loss rate per unit mass $(1.68 \pm 0.10) \times 10^{-11}$\,M$_{\odot}$\,yr$^{-1}$\,M$_{\odot}^{-1}$, translating to a mass-return timescale of $\tau_2\sim5.95 \times 10^{10}$\,yr (Table\,\ref{tab:keyvalues}). 
\begin{table}
	\caption{Key input values adopted for the kinematic model.}
	\begin{tightcenter}
	\begin{tabular}{ l c } 
		\hline
        Parameter & Value\\
        \hline
		Centaurus\,A distance & $3.8$\,Mpc \\
		ISM gas temperature  & $0.65$\,keV \\
		Power-law index for pressure profile  & $-1.50$ \\
        Normalizing radial distance & $54.3$\,arcsec \\
        ISM pressure at normalizing distance & $5.7 \times 10^{-11}$\,dyn\,cm$^{-2}$ \\
        Circular velocity & $250$\,km\,s$^{-1}$ \\
        Observed $M/L_V$ & $3.8$ \\
        BaSTI modelled $M/L_V$ 12\,Gyr population & $2.731$ \\
        BaSTI modelled $M/L_V$ 3\,Gyr population & $1.160$ \\
        Mass-return timescale 12\,Gyr population & $3.333 \times 10^{11}$\,yr \\
        Mass-return timescale 3\,Gyr population & $5.952 \times 10^{10}$\,yr \\
        Fraction 12\,Gyr population & $0.75$ \\
        Fraction 3\,Gyr population & $0.25$ \\
        Entrained fraction & $1.0$ or $0.5$\\
        Jet power & $1.0 \times 10^{43}$\,erg\,s$^{-1}$ \\
        Initial jet speed & $0.667c$ \\
        Jet ratio of specific heats & $13/9$ \\
        Jet viewing angle & $50^{\circ}$ \\
        Solution start & $252$\,pc \\
		\hline
	\end{tabular}
	\end{tightcenter}
	\label{tab:keyvalues}
\end{table}

Direct observational data on $3$\,Gyr-old populations are sparse. At best, the open clusters NGC\,6791 and NGC\,6819, which bound the 3\,Gyr-old population observed in NGC\,5128/Centaurus\,A, can serve as well-studied local comparisons where the stellar death rate can be estimated. For NGC\,6791, aged $\sim8.3$\,Gyr, we find a death rate of 1 star per $\sim7$\,Myr. Given its total mass of $5\times 10^3$\,M$_{\odot}$ \citep{COR17}, $M_{\rm init}\sim1.23$\,M$_{\odot}$ \citep{MIG12} and $M_{\rm fin}\sim0.56$\,M$_{\odot}$ \citep{KAL09}, relying again on equation (\ref{eq:masslossrate}) gives us a rate of $(1/7\times 10^6)\times (0.67/5\times 10^3) \sim1.91\times 10^{-11}$ yr$^{-1}$, or a mass-return timescale of $\tau_{2a}\sim5.22 \times 10^{10}$\,yr. For NGC\,6819, with a population of $\sim2.4$\,Gyr-old stars, we calculate a stellar death rate of 1 per $6$\,Myr. With $M_{\rm tot} = 2.6\times 10^3$\,M$_{\odot}$ \citep{COR17}, $M_{\rm init}\sim1.64$\,M$_{\odot}$ \citep{HAN17} and $M_{\rm fin}\sim0.57$\,M$_{\odot}$ \citep{KAL09}, we have $(1/6\times 10^6)\times (1.07/2.6\times 10^3) \sim6.86 \times 10^{-11}$\,yr$^{-1}$, or a mass-return timescale of $\tau_{2b}\sim1.46\times 10^{10}$\,yr. Then linear interpolation leads to a mass-return timescale for the $3$\,Gyr population of $\tau_{2}\sim1.9 \times 10^{10}$\,yr (the detailed working on those two clusters can be found in Appendix\,\ref{append:ngc6819}). The result is not as near as the modelled value for the $12$\,Gyr population; none the less, it constitutes a valuable check.

Since the modelled values of the mass-loss rate per luminosity and mass-loss rate per mass are in essence the $M/L$ ratios, and represent the $R$-band, we can compare to the BaSTI population synthesis models (see Section\,\ref{sec:fstar}). The modelled $M/L$ for the 12\,Gyr stars gives $2.39$ while the BaSTI $M/L_R$ is $2.48$, and the modelled $M/L$ of the 3\,Gyr population is $0.93$ while the BaSTI $M/L_R = 1.04$. These are in reasonable agreement, and increase the confidence in the modelled mass-return timescales. 
\begin{figure}
	\begin{tightcenter}
	\includegraphics[width=0.53\textwidth]{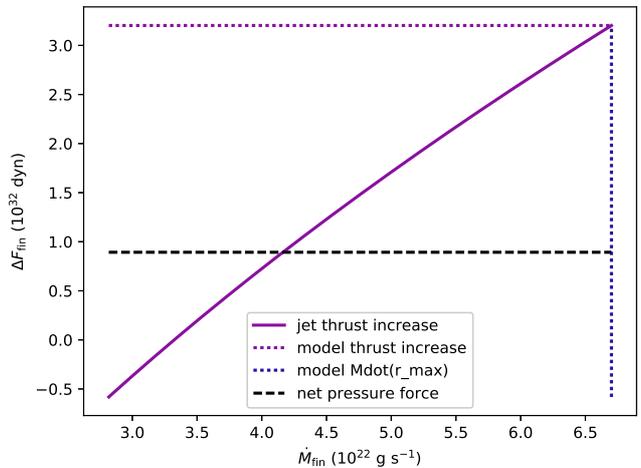}
	\end{tightcenter}
	\caption{Increase in jet thrust versus mass-flow rate, for an initial jet speed of $0.667c$, jet power of $1\times10^{43}$\,erg\,s$^{-1}$ and a jet inclination of $50^{\circ}$. The solid line (purple) shows the increase in jet momentum flux from the initial solution point to the final one. The dashed (black) line signifies the net pressure force on the whole of the modelled jet region. The dotted lines are the model values.}
	\label{fig:fluid-model-fig11}
\end{figure}



\section{Solutions for mass-loaded jet} \label{sec:solutions}

Following the procedure outlined in Section\,\ref{sec:fluid-model}, we solve the energy equation (\ref{eq:power}) for the flow speed as a function of distance from the nucleus $r$. The model parameters used are listed in Table\,\ref{tab:keyvalues}. The initial jet speed determines the value of $\dot{M}$, the flux of rest mass, at the innermost solution point, and then equation (\ref{eq:continuity}) is used to determine $\dot M(r)$ at all other locations.

For reasons discussed in Section\,\ref{sec:intro}, mass is likely deposited in the jet from stellar winds. The material from the surrounding ISM may represent another mass source, but \cite{WYK13, WYK15a} argue that such external entrainment, if occurring in Centaurus\,A, is only a small fraction of the mass injected by stars. Another effect needing consideration before we proceed further is whether all the material lost from stars to the jet is effectively mixed into it, and if not, what is the maximum amount of stellar mass loss allowed by a physically consistent model.

To address the latter point, we plot the increase in momentum flux of the jet (jet thrust) versus mass-flow rate in Fig.\,\ref{fig:fluid-model-fig11}. The solid line here shows the increase in jet momentum flux from the initial solution point to the final one versus the final flux of rest mass through the jet. The lowest value of $\dot M_{\rm fin}$ corresponds to the case that no mass is entrained by the jet in the region modelled. The dashed horizontal line shows the external pressure force on the jet, i.e.\,the net pressure force on the whole of the modelled jet region. The expansion rate of the jet is such that, if no mass were entrained, the jet momentum flux would be inferred to decrease. As outlined in Section\,\ref{sec:fluid-model}, equation (\ref{eq:pforce}) is not used to solve for the jet properties, so that the increase in jet momentum flux provides a constraint on the solution. The jet may experience drag due to its interaction with the surrounding medium or with stationary obstacles within it, in which case the increase in the jet momentum flux could be less than the net pressure force on the jet. However, it is implausible that the momentum flux of the jet increases by more than the net pressure force. From Fig.\,\ref{fig:fluid-model-fig11}, this limits the total mass flux at the final point to $\lesssim 4.2\times10^{22}$\,g\,s$^{-1}$, or $\simeq 40$ per cent of the mass shed by stars within the jet. We are unable to find an acceptable model parameter set that is consistent with equation (\ref{eq:pforce}) if all of the stellar mass loss is entrained. However, the simplified flow model together with the substantial uncertainties in the flow parameters prevent us placing tight constraints on the maximum fraction of the stellar mass loss that can be entrained. Assuming that the drag on the jet is negligible, we have reduced the entrainment rate by a constant factor to match the distribution of the pressure force on the jet. Allowing for the substantial uncertainties here, we have adopted a reduction factor of $0.5$ as representative (see Fig.\,\ref{fig:fluid-model-fig7}).

\begin{figure}
	\begin{tightcenter}
	\includegraphics[width=0.53\textwidth]{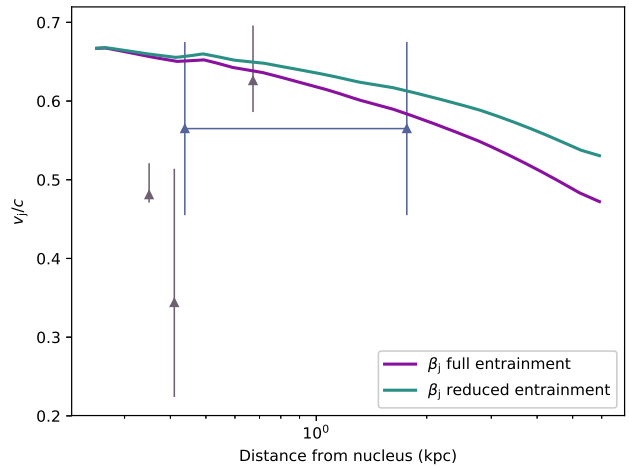}
	\end{tightcenter}
	\caption{Jet velocity versus deprojected distance from the nucleus, for entrained fractions of 1 (purple line) and 0.5 (green line), for an initial jet speed of $0.667c$, jet power of $1\times10^{43}$\,erg\,s$^{-1}$ and an inclination angle of $50^{\circ}$, obtained from solution to equation (\ref{eq:power}). Lower limits on speeds from observations, including error bars, are indicated: at 350, 410 and 672\,pc, the intrinsic component speeds of the A-group knots based on radio data analysis by \citet{GOO10} (triangles, brown), and at 438\,pc to 1.76\,kpc a mean value of the intrinsic component speeds of the A, B and C-group knots based on X-ray data analysis performed by \citet{SNI19} (triangles, cobalt blue).}
	\label{fig:fluid-model-fig5}
\end{figure}

Plots\,\ref{fig:fluid-model-fig5} through\,\ref{fig:fluid-model-fig10} show flow solutions for the parameter set in Table\,\ref{tab:keyvalues}, with entrained fractions of 1 and 0.5. We also discuss the effects of changing the initial jet speed, the jet power, its inclination and the jet width in the remainder of this section.

The criteria for an acceptable model are: (i) the admitted solutions for the jet velocity need to stay above the lower limits from observations; (ii) the run of jet thrust increment should approximately match the run of the net pressure force. The physical assumption is that momentum in the jet is conserved. If the increase in thrust does not match the increase in momentum flux, then there must be forces on the jet that we have not accounted for. That could be the case, but it would be difficult to quantify and is out of the scope of this paper; (iii) the ratio of kinetic to thermal power in the jet should decline with distance from the nucleus. If there was no dissipation, as the pressure decreased, the thermal energy would be transformed into kinetic energy, so that the kinetic energy would keep increasing and the thermal energy decreasing. Dissipation converts kinetic energy to thermal energy, forcing this back in the other direction. This does not necessarily bring the kinetic and thermal fluxes together; however, if the jet is to be decelerated significantly as discussed above, it has to become subsonic.

\begin{figure}
	\begin{tightcenter}
	\includegraphics[width=0.53\textwidth]{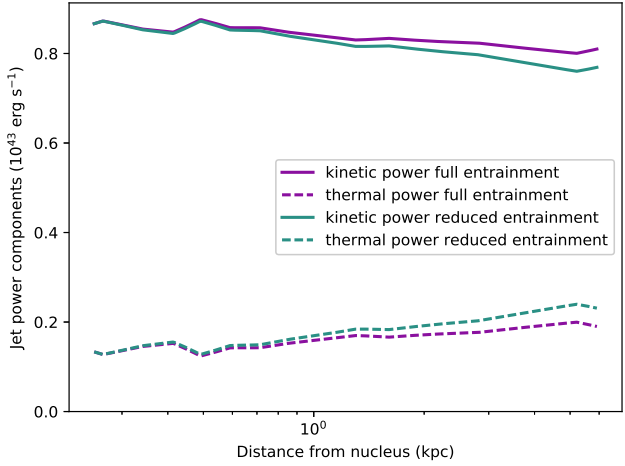}
	\end{tightcenter}
	\caption{Jet power distribution for entrained fractions of 1 (purple) and 0.5 (green), for an initial jet speed of $0.667c$, jet power of $1\times10^{43}$\,erg\,s$^{-1}$ and a jet inclination of $50^{\circ}$, as a function of deprojected distance from the nucleus. The solid lines indicate the kinetic power, the dashed lines the thermal power ($A \Gamma_{\rm j}^2 v_{\rm j} (w - \rho_{\rm j} c^2)$, see also equation ({\ref{eq:power}})).}
	\label{fig:fluid-model-fig6}
\end{figure}

\begin{figure}
	\begin{tightcenter}
	\includegraphics[width=0.53\textwidth]{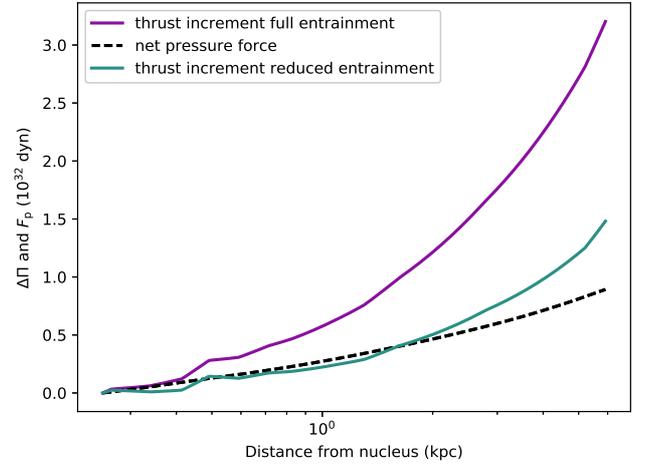}
	\end{tightcenter}
	\caption{Momentum flux increment (solid line) and net pressure force (dashed line, black), for entrained fractions of 1 (purple) and 0.5 (green), for an initial jet speed of $0.667c$, jet power of $1\times10^{43}$\,erg\,s$^{-1}$ and a jet inclination of $50^{\circ}$, as a function of deprojected distance from the nucleus. Neither the increase in the momentum flux, nor the net pressure force is used in the solution.}
	\label{fig:fluid-model-fig7}
\end{figure}

Plotted in Fig.\,\ref{fig:fluid-model-fig5} is $\beta_{\rm j}$, the ratio of the jet speed $v_{\rm j}$ to the speed of light, against the physical distance from the nucleus. To connect to observations, we include the intrinsic speeds from the proper motion measurements: radio proper motions of $\beta_{\rm j,app} = 0.534^{+0.06}_{-0.02}$, $0.338^{+0.22}_{-0.15}$ and $0.802^{+0.15}_{-0.09}$ \citep{GOO10} give through the Doppler formula $\beta_{\rm j} = \beta_{\rm j,app} / ({\rm sin}\, \theta + \beta_{\rm j,app}\, {\rm cos}\, \theta)$ intrinsic speeds of $\beta_{\rm j} = 0.481^{+0.04}_{-0.01}$, $0.344^{+0.17}_{-0.12}$ and $0.626^{+0.07}_{-0.04}$, respectively, and the X-ray proper motion of $\beta_{\rm j,app} = 0.68^{+0.20}_{-0.20}$ \citep{SNI19} leads to $\beta_{\rm j} = 0.565^{+0.11}_{-0.11}$. These are treated as lower limits to the bulk-flow speed. Keeping other things equal, increasing the initial jet bulk-flow speed causes the thermal power to rise at the expense of kinetic power (Fig.\,\ref{fig:fluid-model-fig6}). With a higher speed, the mass flow must be lower to satisfy equation (\ref{eq:power}). A higher velocity better fits both the thermal power and momentum-pressure force gauges of the model; models with initial jet velocity of less than $0.65c$ are difficult to sustain (and at any rate, they are barely supported by observations) but models with initial velocity $\ge0.65c$ work well. At the higher end, the limit for a reasonable model is $0.70c$. A velocity drop to $\sim0.47c$ towards the end of the jet with the full entrainment and to $\sim0.52c$ in the entrainment reduced by $50$ per cent is representative of most of the runs. A smaller versus larger inclination angle causes the terminal velocity to respectively increase and decrease.

\begin{figure}
	\begin{tightcenter}
	\includegraphics[width=0.50\textwidth]{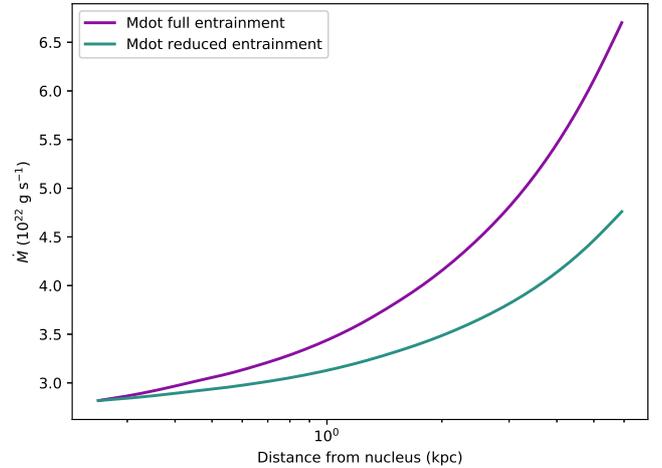}
	\end{tightcenter}
	\caption{Mass-flow rate for entrained fractions of 1 (purple line) and 0.5 (green line), for an initial jet speed of $0.667c$, jet power of $1\times10^{43}$\,erg\,s$^{-1}$ and a jet inclination of $50^{\circ}$, as a function of deprojected distance from the nucleus.}
	\label{fig:fluid-model-fig8}
\end{figure}

Fig.\,\ref{fig:fluid-model-fig6} compares the kinetic and thermal components of the jet power, given by the two terms on the right-hand side of equation (\ref{eq:power}). It is clear from the figure that the kinetic power is dominant throughout the jet. The kinetic power is even more dominant for higher jet powers and for smaller inclination angles. The ratio of initial velocity to terminal velocity in Fig.\,\ref{fig:fluid-model-fig5}, expressed in terms of $(\Gamma_{\rm j} - 1)$, shows a decrease in $(\Gamma_{\rm j} - 1)$ of $0.342/0.133 \sim2.57$ (full entrainment) and $0.342/0.171 \sim2.0$ (reduced entrainment). The ratio of final $\dot M$ over initial $\dot M$ in Fig.\,\ref{fig:fluid-model-fig8} represents an increase of $6.7/2.8\sim2.39$ (full entrainment) and of $4.8/2.8\sim1.71$ (reduced entrainment). Those values are sufficiently close to one another and support the assumption that the jet power $P_{\rm j}$ does not significantly change and the thermal part of the proper enthalpy density $h$ can be neglected. 

Fig.\,\ref{fig:fluid-model-fig7} displays the thrust increment, or increase in jet momentum flux from the initial solution point $\Delta \Pi$ and the cumulative net pressure force (defined on the right in equation (\ref{eq:pforce})). The net pressure force acting on the jet can increase its momentum flux. The net pressure force is fully determined by the pressure profile and jet area, so it does not depend on other jet properties. Changes in the initial speed do affect the thrust increment in the sense that higher speeds give higher $\Delta \Pi$. Both the momentum flux and the net pressure force are affected by the pressure: increasing the pressure will generally reduce the momentum flux, while increasing the net pressure force. Fig.\,\ref{fig:fluid-model-fig11} lead to the conclusion that the best choice of injection rate is about $40$ per cent of the total mass-loss rate from the stars; however, this only provided the comparison at one location. From Fig.\,\ref{fig:fluid-model-fig7}, we can conclude it should be somewhat greater. Appendix\,\ref{append:add-plots} additionally shows the behaviour of the jet momentum in case of zero mass entrainment (Fig.\,\ref{fig:fluid-model-figC2}): the apparent drop in jet momentum strongly suggests that the jet entrains mass. The jet velocity diminishes only marginally for zero mass entrainment (Fig.\,\ref{fig:fluid-model-figC1}), corroborating this interpretation. See the appendix for further discussion.

In Fig.\,\ref{fig:fluid-model-fig8}, the run of the mass-flow rate $\dot M$ (obtained from equation (\ref{eq:eq11})) is plotted versus distance from the nucleus. $\dot M$ increases monotonically outwards. As obvious from equation (\ref{eq:power}), if $h$ is negligible, $P_{\rm j}\propto (\Gamma_{\rm j}-1) \dot M$. Smaller inclination angles $\theta$ reduce the deprojected volume
of the jet, hence lowering $\dot{M}$ at a fixed projected radius and vice versa. The resulting value for $\theta=50^{\circ}$ is $\dot M \sim6.7\times10^{22}$\,g\,s$^{-1}$ ($\sim1.1 \times 10^{-3}$\,M$_{\odot}$\,yr$^{-1}$) and $\sim4.8\times10^{22}$\,g\,s$^{-1}$ ($\sim7.6 \times 10^{-4}$\,M$_{\odot}$\,yr$^{-1}$) for respectively the full and the reduced entrainment. This is within an order of magnitude of the value of $\dot M \sim 1.4\times10^{23}$\,g\,s$^{-1}$ ($\sim2.3 \times 10^{-3}$\,M$_{\odot}$\,yr$^{-1}$) derived by \cite{WYK15a}. This rate of mass injection is also sufficient to slow down Centaurus\,A's $\sim1\times10^{43}$\,erg\,s$^{-1}$ jet (as already demonstrated by \citealp{WYK15a}), but generally not FR\,II jets (e.g.\,\citealp{KOM94, HUB06, PER14a, PER14b}). 

\begin{figure}
	\begin{tightcenter}
	\includegraphics[width=0.53\textwidth]{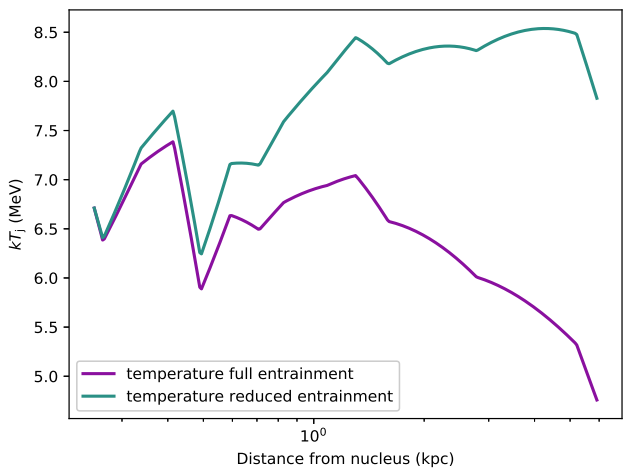}
	\end{tightcenter}
	\caption{Jet `temperature' for entrained fractions of 1 (purple line) and 0.5 (green line), for an initial jet speed of $0.667c$, jet power of $1\times10^{43}$\,erg\,s$^{-1}$ and a viewing angle of $50^{\circ}$, as a function of deprojected distance from the nucleus.}
	\label{fig:fluid-model-fig9}
\end{figure}
\begin{figure}
	\begin{tightcenter}
	\includegraphics[width=0.53\textwidth]{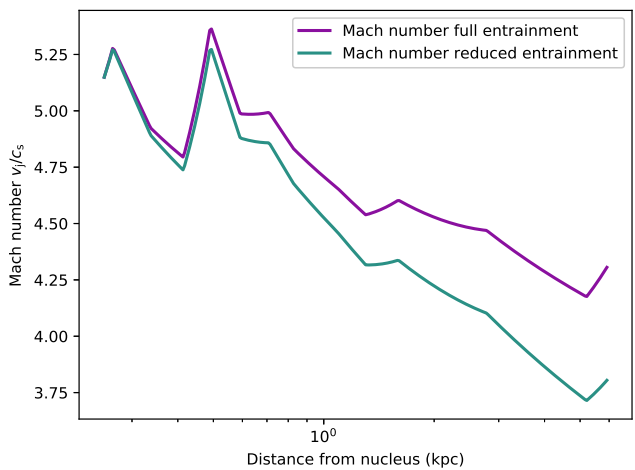}
	\end{tightcenter}
	\caption{Sonic Mach number for entrained fractions of 1 (purple line) and 0.5 (green line), for an initial jet speed of $0.667c$, jet power of $1\times10^{43}$\,erg\,s$^{-1}$ and a jet inclination of $50^{\circ}$, as a function of deprojected distance from the nucleus.}
	\label{fig:fluid-model-fig10}
\end{figure}
Fig.\,\ref{fig:fluid-model-fig9} shows the effective temperature of the jet, obtained from the pressure and jet density, using the ideal gas law, $p = \rho_{\rm j} kT / (\mu m_{\rm H})$, where the pressure is as defined in equation (\ref{eq:pressure}) and the proper density of the jet is obtained from equation (\ref{eq:mdot}). The jet plasma is unlikely to be in thermal equilibrium. Nevertheless, the value of $kT$ obtained this way should be representative of the typical particle energies. The temperature profile in the figure is ranging between $\sim4.7$ and $8.6$\,MeV, which supports our choice for the input ratio of specific heats (relativistic electrons, subrelativistic protons). Within the model, relativistic protons are not required. Increasing the rate of entrainment raises $\dot{M}$ along the jet. With all other parameters in the energy equation (\ref{eq:power}) fixed, this reduces the solution for the jet speed and, combined with the direct effect of lower $\dot M$ in equation (\ref{eq:mdot}), increases the value of the jet density. Since the pressure is fixed, the temperature must drop with decreasing entrainment fraction; the curve of full entrainment in Fig.\,\ref{fig:fluid-model-fig9} demonstrates this behaviour, and the curve of $50$ per cent entrainment shows a radially raising trend. The glitches are more evident in this plot due to the small temperature range covered.

Fig.\,\ref{fig:fluid-model-fig10} displays the internal sonic Mach number of the jet, defined here as \begin{equation}
\mathcal{M} = \frac{v_{\rm j}}{c_{\rm s}}\,,
\end{equation}
with $c_{\rm s}$ the internal sound speed. The expression for the sound speed for a relativistic fluid depends on the proper pressure and density of the jet, and does not require the temperature to be well defined:
\begin{equation}
\Big(\frac{c_{\rm s}}{c}\Big)^2 = \frac{\phi}{1 + \phi / (\gamma - 1)}\,,
\end{equation}
where the parameter $\phi = \gamma p / (\rho c^2)$, and the ratio of specific heats $\gamma$ is the input value (Section\,\ref{sec:methodology}) $\gamma=13/9$. At a temperature of $6.5$\,MeV (Fig.\,\ref{fig:fluid-model-fig9}), the sound speed in the jet is about $0.14c$, which means that pressure changes are communicated very rapidly across the jet. As illustrated in Fig.\,\ref{fig:fluid-model-fig10}, the Mach number spans a range of $\mathcal{M}=3.6-5.3$; it declines as more material is entrained in the jet. The flow is not subsonic by the end of the visible X-ray jet: this is consistent with the thermal power remaining below the kinetic power by that end (Fig.\,\ref{fig:fluid-model-fig6}). If the internal density is much smaller than the external density, a Mach number 3.6 shock would not necessarily be visible in X-rays and so not disagree with the {\it Chandra} observations of the jet. On the other hand, the `flare point' indicated in figures 2 and 3 of \cite{HAR06}, seen in X-ray/infrared/radio bands just before the X-ray jet tapers off, could represent a shock.


\section{Discussion} \label{sec:discussion}

While our model draws analogies to seminal works such as those by \cite{BLA74}, \cite{BIC94}, \cite{KOM94}, \cite{BOW96} and \cite{LAI02a}, there are a number of differences. The work by \cite{LAI02a} in particular, on the FR\,I source 3C\,31, differs from our method as follows.\\
(i) In 3C\,31, pressure equilibrium was not assumed everywhere. The jet was allowed to be out of equilibrium at the start of the flaring region (rapid expansion). Pressure equilibrium at large distances was required, at least in the `reference' model. Such a model may not be appropriate for Centaurus\,A, which does not show heavy flaring.\\
(ii) The mass injection was not set {\it a priori} in 3C\,31, since external entrainment was included. For Centaurus\,A, we consider internal stellar mass loss to be more important.\\
(iii) The area and the angle to the line of sight were fixed, since these were determined relatively well by the kinematic model.\\
(iv) Velocity versus distance were at least constrained (to within variation across the jet width).\\
(v) The equation of state was relativistic.\\
(vi) The jet flow was not required to be irrotational and one-dimensional.\\
(vii) The work found lower (relativistic) Mach numbers than the Mach numbers in our work on Centaurus\,A.

Among the strengths of our model, it is configured to match a number of observed properties of the jet. In particular, the power, initial speed and profile of the jet are constrained by observations, and the pressure profile of the jet is matched to that of the adjacent ISM, as required for a steady, pressure-confined flow. The model is also consistent with an entrainment rate that is comparable to the estimated stellar mass-loss rate within the jet. This means that entraining the stellar mass loss would produce a dissipation rate in the jet close to that required to account for the observed pressure profile and width of the jet. These are testing constraints, since the mass-loss rate could have been orders of magnitude different.

Among the less fortunate features of the model, it is unsatisfactory that we need to reduce the entrainment rate by a factor of roughly 2 in order to get models that conserve momentum. There are two possible causes for this. The first is that the jet does not entrain all of the stellar mass loss. In that case, it is surprising that the entrained fraction is so close to unity. In particular, the results of \cite{COO09} suggest that it should be much smaller. The second possibility is that the approximations of our model are too crude. This is hard to evaluate without more sophisticated modelling, but it is certainly possible that a more realistic flow model might account for the apparent discrepancy of a factor of $2$. The steady flow model remains promising, but a more realistic implementation is required to determine if it can fully account for the properties of the Centaurus\,A jet. 

Some lesser issues are considered below.

We could ask at what level a potential ISM stratification will affect the deduced pressure, as the ISM gas is likely to deviate significantly from spherical symmetry (e.g.\,\citealp{CRO09}). Our regions cross sharp features in the ISM gas; hence, there must be multiple temperatures and partial covering absorption that we do not account for. NGC\,5128 has a complex history. Then, while all components must be in the same potential, they may have different assembly histories and therefore different kinematics (e.g.\,\citealp{PEN04b}), which makes comparisons difficult. 

It is not obvious that core jets (VLBI-scale jets embedded in VLA-scale cores) are in pressure equilibrium with the surrounding ISM. In fact, it is more likely that this is not the case (see e.g.\,\citealp{BEG84} and \citealp{BIC94}) and that they are either in free expansion, magnetically self-confined, or confined by a slower-moving wind or a galactic fountain. Whatever the mechanism, this is not expected to affect our pressure analysis because our lower sector border is placed at $\sim190$\,pc (projected) from the nucleus where the jet width has already expanded to $\sim70$\,pc, and so is likely confined by the ISM (\citealp{BIC94} quotes a limiting jet diameter of $\sim30$\,pc and distance from the nucleus of the order of $300$\,pc). The recently discovered galactic fountain narrowly following the jet to at least 180\,pc projected downstream \citep{ISR17}, with pressure of about $8\times10^{-11}$\,dyn\,cm$^{-2}$, could be responsible for a moderate extra external pressure on these scales. Since, in addition, this concerns a relatively small portion of our designed sector area, we are confident that the presence of the galactic fountain does not significantly impact on our deductions.

We used the disc rotation speed of \cite{GRA79} to estimate the gravitating matter density in NGC\,5128, since there is evidence of significant departures from full hydrostatic equilibrium in the hot ISM \citep{KRA09}. Evidence against full hydrostatic equilibrium also exists for some other systems \citep{ASC06, VAZ18}.

Using a magnetic equation of state in the fluid model would increase the ratio of specific heats and hence increase the difference between the thermal and kinetic power and also between the momentum flux increment and net pressure force, and so make the model perform worse. However, at kpc jet scales, the effect of magnetic fields is not dominant (e.g.\,\citealp{SIK05, KOM07}) and we feel justified in neglecting it.

That local dissipation (and particle acceleration) is required to explain the observations of jets has been recognised for some time (e.g.\,\citealp{FER79, BIC82, BEG84}). However, it affects internal pressure or momentum only at a very low level.

As also pointed out by \cite{BLA74} and \cite{POR15}, and shown above (Section\,\ref{sec:fluid-model}), it seems unlikely that all instabilities could be suppressed; however, they may not grow sufficiently to alter entirely the nature of the flow and disrupt the jet. Interestingly, the limit on $\dot{M}$ applies to the entrainment via stellar mass loss as well as the entrainment from the jet boundary. Specifically, \cite {WYK13} calculated the mean inflow rate along the boundary to be $\sim3.0\times10^{21}$\,g\,s$^{-1}$ ($\sim4.7 \times 10^{-5}$\,M$_{\odot}$\,yr$^{-1})$ for the jet within 3\,kpc from the nucleus, approaching an order of magnitude smaller than the rate required by the models presented here.

\cite{WYK13, WYK15a} did not investigate the interaction of the jet with clouds potentially drifting into its path. Our modelling here shows that the properties of the jet can be accounted for reasonably well if the jet entrains a substantial fraction of the mass shed into it by stars (and since the model is approximate, possibly all of it). The work done on a slowly-moving obstacle in the jet is negligible, so if the brightest knots are due to interaction with molecular clouds, they will not alter the jet power appreciably, but they could reduce the jet momentum flux, in which case we have overestimated the amount of mass entrained by the jet. Except in the unlikely event that the jet is less effective at entraining mass from stellar winds than from molecular clouds, any molecular clouds cannot have a great impact compared to the mass shed by the stars. We have no evidence for clouds in the vicinity of the Centaurus\,A X-ray jet; the presence of clouds is merely demonstrated for the `middle regions' (e.g.\,\citealp{SAL17} and references therein), several kpc beyond the extent of the X-ray jet.

With the power, pressure and cross-sectional area of the jet fixed by observations, specifying the speed of the jet at the initial point in the power equation (\ref{eq:power}) determines the kinetic power and the flux of rest mass. Despite the uncertainties in the jet parameters, the large initial disparity between the kinetic and thermal powers of the jet (Fig.\,\ref{fig:fluid-model-fig6}) is hard to avoid. Changing this significantly would require a substantially smaller jet power or substantially greater initial speed. The initial dominance of the kinetic power is the main reason for the high jet momentum flux and Mach number throughout the modelled region. The relatively high final internal Mach number of the jet, about $4$, could present a challenge to understanding the flow of the jet beyond the modelled region.

The greatest simplification in the fluid model is the assumption of a constant jet speed across the width of the jet. We know from modelling of other sources (e.g.\,\citealp{GHI05, GOP07, LAI14, SOB17}) that this is not likely to be accurate. As discussed in Section\,\ref{sec:fluid-model}, the conical tip on the jet places some constraint on the velocity difference between the spine and sheath. It probably is not large. On the other hand, we cannot say that it is insufficient to account for the modest discrepancy in the entrained mass. At any rate, concentrating the energy around the jet spine and the momentum in the sheath, i.e.\,designing a multizone analytical model, would no longer allow the model to be fully determined by the data. Then numerical HD or MHD simulations would be more appropriate. The model accounts reasonably well for the observed properties, but requires only about half of the stellar mass loss to be entrained. This suggests that a more accurate flow model that entrains all of the stellar mass loss might well be fully consistent with the observed properties of the jet.



\section{Summary} \label{sec:summary}

We have presented results for a steady, one-dimensional hydrodynamical model for flow in the jet of Centaurus\,A. The pressure profile of the jet is constrained by pressures in the ISM of NGC\,5128 determined from $\sim260$\,ks of new and archival {\it Chandra}/ACIS observations in regions adjacent to the jet. The width of the jet is also determined from radio and X-ray data. The flow model conserves particle number and energy, while conservation of momentum is used to provide an additional constraint. This tests the scenario of decelerating flows via stellar-mass entrainment pertaining to FR\,I jets. The main results are as follows.

(1) The pressure profile of the host galaxy atmosphere adjacent to the jet is adequately modelled as a power law of the form $p(r) \propto r^{-1.5}$, decreasing from $\sim1.4\times10^{-10}$ to $\sim6.2\times10^{-12}$\,dyn\,cm$^{-2}$ between $0.2$ and $5.5$\,kpc deprojected distance from the nucleus. We find an internal jet density of about $3 \times 10^{-4}$ of the density of the surrounding ISM.

(2) Based on mass-to-light ratios of the 12 and 3 Gyr stellar populations in NGC\,5128/Centaurus\,A, we estimate the fraction of gravitating mass in stars to be $\sim0.54$. Relying on stellar evolution models, we compute a mass-return timescale of about $3.33 \times 10^{11}$\,yr for the NGC\,5128's $\sim12$\,Gyr-old $(Z\sim0.004)$ population and $\sim5.95 \times 10^{10}$\,yr for its $\sim3$\,Gyr $(Z\sim0.008)$ population; this agrees with mass-return timescales from direct observations of similar stars in nearby stellar clusters.

(3) The simple fluid model of the jet whose solutions are irrotational and anisentropic, and ensure
conservation of particles and energy, captures the gross features well. For this model, not all mass
lost by stars into the jet is entrained/well-mixed; the entrained fraction is $0.5$, corresponding to $\sim4.8\times10^{22}$\,g\,s$^{-1}$, or $\sim7.6 \times 10^{-4}$\,M$_{\odot}$\,yr$^{-1}$. A more accurate hydrodynamical model may allow all of the stellar mass loss to be entrained.

(4) The jet is best modelled as an initially moderately relativistic flow with intrinsic velocity $\sim0.67c$, declining to $\sim0.52c$ by the end of the X-ray jet, with a jet power of $\sim1.0 \times 10^{43}$\,erg\,s$^{-1}$ and inclination $\sim 50^{\circ}$. The temperature profile of the jet varies in the range $\sim8.6-4.7$\,MeV, and the sonic Mach number in the range $\sim5.3-3.6$. The injection of stellar wind material appears to be able to account for virtually all the internal dissipation.


\section*{Acknowledgements}

We acknowledge helpful conversations with Robert Laing, Chris O'Dea, Nicky Brassington and Ken Freeman. We would also like to thank the referee for a thoughtful report. SW thanks the Harvard-Smithsonian CfA for a research fellowship. PEJN and RPK were supported in part by NASA contract NAS8-03060. MJH acknowledges support from the UK's Science and Technology Facilities Council (grant number ST/R000905/1). TWJ acknowledges support from the US NSF grant AST1714205. Support for this work was provided by the National Aeronautics and Space Administration through {\it Chandra} Award Number G07-18104X issued by the {\it Chandra} X-ray Center, which is operated by the Smithsonian Astrophysical Observatory for and on behalf of the National Aeronautics Space Administration under contract NAS8-03060.




\appendix

\section{Spectral analysis details and ISM deprojection results} \label{append:deproj}

Here, we present an overview of the spectral deprojection data used in Section\,\ref{sec:deprojection}. Tables\,\ref{table:east} and \ref{table:west} include the inner annulus diameter $d_{\rm in}$, the outer annulus diameter $d_{\rm out}$, the H\,{\small I} column density $N_{\rm H}$, and the deprojected temperature $kT$, electron density $n_{\rm e}$ and pressure $p$. The H\,{\small I} column density was freed for the regions within the dust lane, while for all other regions this was frozen to the Galactic column density. 

The abundances of Centaurus\,A's ISM are also available from the model fits. The results are presented in Table\,\ref{tab:abundances}.
\begin{table*}
	\caption{Inner and outer sector borders (distance from nucleus, projected values), column densities, and the best-fitting deprojected temperature, electron density and pressure of the thermal emission of Centaurus\,A's ISM near the main jet. The error bars indicate 90 per cent confidence intervals. Results for the eastern sector.}
	\label{table:east}
	\begin{tightcenter}
	\begin{tabular}{ c c c c c c }
		\hline
		$d_{\rm in}$ & $d_{\rm out}$ & $N_{\rm H}$           & $kT$  & $n_{e}$      & $p$ \\ 
		(kpc)        & (kpc)         &($10^{22}$\,cm$^{-2}$) & (keV) & ($10^{-3}$\,cm$^{-3}$) & ( $10^{-11}$ dyn\,cm$^{-2}$)\\ 
		\hline
		$0.19$ & $0.86$ & $0.59\substack{+0.06\\-0.06}$ & $0.95\substack{+0.06\\-0.06}$ 
			&  $45.94\substack{+2.80\\-2.68}$           & $13.42\substack{+1.76\\-1.60}$ \\
		$0.86$ & $1.48$ & $1.06\substack{+0.23\\-0.17}$ & $1.09\substack{+0.21\\-0.15}$ 
			&  $25.64\substack{+1.63\\-3.44}$ & $8.64\substack{+2.35\\-2.16}$ \\
		$1.48$ & $2.07$ & $0.084$             & $0.74\substack{+0.02\\-0.02}$ &  $14.66\substack{+0.32\\-0.30}$ & $3.34\substack{+0.18\\-0.17}$ \\
		$2.07$ & $2.60$ & $0.084$             & $0.79\substack{+0.18\\-0.06}$ &  $7.43\substack{+0.34\\-0.27}$ & $1.81\substack{+0.52\\-0.21}$ \\
		$2.60$ & $3.16$ & $0.084$ & $0.73\substack{+0.06\\-0.06}$ &  $6.18\substack{+0.29\\-0.29}$ 
			& $1.40\substack{+0.19\\-0.17}$ \\
		$3.16$ & $3.78$ & $0.084$ & $0.64\substack{+0.05\\-0.05}$ &  $5.33\substack{+0.24\\-0.24}$ 
			& $1.05\substack{+0.13\\-0.13}$ \\
		$3.78$ & $4.57$ & $0.084$ & $0.68\substack{+0.09\\-0.09}$ &  $2.67\substack{+0.23\\-0.24}$ 
			& $0.56\substack{+0.13\\-0.11}$ \\
		$4.57$ & $5.53$ & $0.084$ & $0.63\substack{+0.04\\-0.03}$ &  $3.13\substack{+0.17\\-0.25}$ 
			& $0.61\substack{+0.07\\-0.07}$ \\
		\hline
		\end{tabular}
	\end{tightcenter}
    \label{tab:params-east}
    {${}$ Note. Foreground contamination: $kT = 0.213$\,keV, Abund $= 1$ (fixed), Norm $= 1.41 \times 10^{-4}$. $\beta$ model: $kT = 0.628$\,keV (uses the outermost shell's temperature), Abund $= 0.3$ (fixed), $N_{\rm H} = 0.084 \times 10^{-22}$\,cm$^{-2}$ (fixed), Norm = $4.39 \times 10^{-4}$ (uses the outermost shell's normalization).} 
\end{table*}
\begin{table*}
	\caption{Same as Table\,\ref{tab:params-east} but for the western sector.}
	\label{table:west}
	\begin{tightcenter}
	\begin{tabular}{ c c c c c c }
		\hline
		$d_{\rm in}$ & $d_{\rm out}$ & $N_{\rm H}$           & $kT$  & $n_{e}$               & $p$ \\ 
		(kpc)        & (kpc)         &($10^{22}$\,cm$^{-2}$) & (keV) & ($10^{-3}$\,cm$^{-3}$)& ( $10^{-11}$ dyn\,cm$^{-2}$)\\ 		
		\hline
		$0.19$ & $0.86$ & $0.75\substack{+0.07\\-0.07}$ & $1.01\substack{+0.10\\-0.10}$ 
			& $45.39\substack{+2.87\\-2.63}$ & $14.22\substack{+2.41\\-2.09}$ \\
		$0.86$ & $1.48$ & $0.16\substack{+0.04\\-0.04}$ & $0.78\substack{+0.04\\-0.05}$ 
			& $16.10\substack{+0.96\\-0.90}$ & $3.88\substack{+0.46\\-0.44}$ \\
		$1.48$ & $2.07$ & $0.084$ & $0.77\substack{+0.04\\-0.05}$ & $9.02\substack{+0.33\\-0.32}$ 
			& $2.15\substack{+0.21\\-0.20}$ \\
		$2.07$ & $2.60$ & $0.084$ & $0.53\substack{+0.10\\-0.10}$ & $4.64\substack{+0.40\\-0.42}$ 
			& $0.76\substack{+0.22\\-0.20}$ \\
		$2.60$ & $3.16$ & $0.084$ & $0.42\substack{+0.11\\-0.06}$ & $4.34\substack{+0.38\\-0.34}$ 
			& $0.56\substack{+0.21\\-0.12}$ \\
		$3.16$ & $3.78$ & $0.084$ & $0.64\substack{+0.05\\-0.07}$ & $4.35\substack{+0.21\\-0.21}$ 
			& $0.86\substack{+0.12\\-0.13}$ \\
		$3.78$ & $4.57$ & $0.084$ & $0.64\substack{+0.15\\-0.08}$ & $3.03\substack{+0.18\\-0.18}$ 
			& $0.60\substack{+0.19\\-0.11}$ \\
		$4.57$ & $5.53$ & $0.084$ & $0.64\substack{+0.02\\-0.02}$ & $3.18\substack{+0.06\\-0.04}$ 
			& $0.62\substack{+0.03\\-0.03}$ \\
		\hline
		\end{tabular}
	\end{tightcenter}
	\label{tab:params-west}
\end{table*}
\begin{table}
	\caption{Elemental abundances for the eastern and western sectors from spectral deprojection fits. Only values for the regions outside the dust lane are considered. $\rm [\alpha/Fe]$ is defined as $\rm log (\alpha/Fe)_{ISM}$. The error bars indicate 90 per cent confidence intervals.}
	\label{table:abundance}
	\begin{tightcenter}
	\begin{tabular}{ l c c }
		\hline
		Element ratio & Eastern sector & Western sector \\ 
		\hline
		$\rm[O/Fe]$ & $0.74\substack{+0.05\\-0.05}$ & $1.06\substack{+0.05\\-0.05}$ \\
		$\rm[Ne/Fe]$ & $0.44\substack{+0.08\\-0.09}$ & $0.45\substack{+0.08\\-0.11}$ \\
		$\rm[Mg/Fe]$ & $0.39\substack{+0.04\\-0.05}$ & $0.46\substack{+0.06\\-0.07}$ \\
		$\rm[Si/Fe]$ & $0.41\substack{+0.05\\-0.05}$ & $0.41\substack{+0.06\\-0.07}$ \\
		\hline
	\end{tabular}
	\end{tightcenter}
	\label{tab:abundances}
\end{table}  

\begin{table}
	\caption{Locations along the length of the X-ray jet where the jet angular width was measured. The distances from nucleus were selected such that they sample areas where the jet width changes sharply, and to also sufficiently cover the entire jet. The error bars reflect the $0.5$ {\it Chandra} pixel uncertainty.}
	\begin{tightcenter}
	\begin{tabular}{ c c }
		\hline
		Distance         & Opening        \\
	    from nucleus     & angle          \\
		(kpc, projected) & ($^{\circ}$)   \\
		\hline
		$0.136$          & $20.4\substack{+1.8\\-1.9}$       \\
		$0.204$          & $16.5\substack{+1.0\\-1.2}$       \\
		$0.258$          & $16.9\substack{+1.0\\-1.0}$       \\
		$0.317$          & $16.7\substack{+0.8\\-0.8}$       \\
		$0.376$          & $14.4\substack{+0.7\\-0.6}$       \\
		$0.453$          & $15.0\substack{+0.6\\-0.6}$       \\
		$0.543$          & $14.5\substack{+0.5\\-0.5}$       \\
		$0.634$          & $14.6\substack{+0.4\\-0.4}$       \\
		$0.833$          & $14.5\substack{+0.3\\-0.3}$       \\
		$0.996$          & $14.5\substack{+0.3\\-0.2}$       \\
		$1.222$          & $13.9\substack{+0.2\\-0.2}$       \\
		$1.584$          & $13.7\substack{+0.1\\-0.2}$       \\
		$2.127$          & $13.4\substack{+0.1\\-0.2}$       \\
		$3.983$          & $13.6\substack{+0.1\\-0.1}$       \\
		$4.526$          & \hspace{5pt}$13.1\substack{+0.1\\-0.1} ^a$   \\
		\hline
	\end{tabular}
	\end{tightcenter}
	\label{tab:locations}
	{${}^{a}$ Measurement from VLA observations.}
\end{table}

\clearpage
\newpage


\section{NGC\,6791 and NGC\,6819 analysis} \label{append:ngc6819}

NGC\,6791 is a metal-rich open cluster of age $\sim8.3$\,Gyr with total mass of about $5000$\,M$_\odot$ \citep{COR17}. There are roughly $30$ stars in or brighter than the red clump \citep{LOO08}, a criterion that includes the upper-RGB, core-helium-burning and AGB phases of evolution. The initial masses of evolved stars in NGC\,6791 are $M_{\rm init} \sim 1.23$\,M$_\odot$, reducing to $\sim1.14$\,M$_\odot$ by the red clump itself \citep{MIG12}. Stellar evolutionary tracks with mass $M_{\rm init} = 1.2$\,M$_\odot$, metallicity $Z = 0.04$ and helium fraction $Y = 0.026$ were extracted from \cite{BER08}, indicating that these phases of evolution last $\sim210$\,Myr. Hence, the stellar death rate in NGC\,6791 is 1 star per $\sim7$\,Myr, or $\sim2.9\times 10^{-11}$\,yr$^{-1}$\,M$_\odot^{-1}$. White dwarf masses in the cluster are $\sim0.56$\,M$_\odot$ \citep{KAL09}; hence, stars return $\sim0.67$\,M$_\odot$ to the ISM each, giving a mass-loss rate of $\sim1.91\times 10^{-11}$ yr$^{-1}$, or a mass-return timescale of $\sim5.22 \times 10^{10}$\,yr.

NGC\,6819 is a solar-metallicity cluster of age $\sim2.4$\,Gyr with total mass of approximately $2600$\,M$_\odot$ \citep{COR17}. There are roughly $39$ stars in or brighter than the RGB bump (and this includes $\sim9$ that have experienced some form of `non-standard evolution'), which began with $M_{\rm init} \sim 1.64$\,M$_\odot$, with negligible mass lost before the red clump \citep{HAN17}. A stellar evolution track with $M_{\rm init} = 1.6$\,M$_\odot$, $Z = 0.017$ and $Y = 0.023$ \citep{BER08} indicates that stars first attaining the luminosity of the bottom of the RGB bump have $\sim234$\,Myr left to live. This yields a stellar death rate in NGC\,6819 of 1 star per $6$\,Myr, or $\sim6.4 \times 10^{-11}$\,yr$^{-1}$\,M$_\odot^{-1}$. The cluster's white dwarfs are $\sim0.57$\,M$_\odot$ in mass \citep{KAL09}; hence, each star returns $\sim1.07$\,M$_\odot$ to the ISM, giving a mass-loss rate of $\sim6.86 \times 10^{-11}$\,yr$^{-1}$, or a mass-return timescale of $\sim1.46\times 10^{10}$\,yr.

The anticipated uncertainties in these timescales are $\sim35$ per cent from Poisson noise, lack of completeness, and inclusion of non-members in the star counts; $\sim20$ per cent in the calculation of evolutionary timescales, due to simplifications made in the input parameters, applicability of the individual tracks used and uncertainties from the treatment of late-stage stellar evolution; and $\sim5$ per cent in the total mass lost by each star giving a total uncertainty in the final timescale of $\sim40$ per cent. Hence, a reasonable estimate for the mass-return timescale for the 3\,Gyr-old population of Centaurus\,A would be $\sim(1.9 \pm 0.5) \times 10^{10}$\,yr. We add that metal-poor stars as we consider in Centaurus\,A will evolve faster. The mass-return timescale will be increased by about $10$ per cent for half of solar metallicity. 



\section{Solutions for unloaded jet} \label{append:add-plots}

\begin{figure}
	\begin{tightcenter}
	\includegraphics[width=0.53\textwidth]{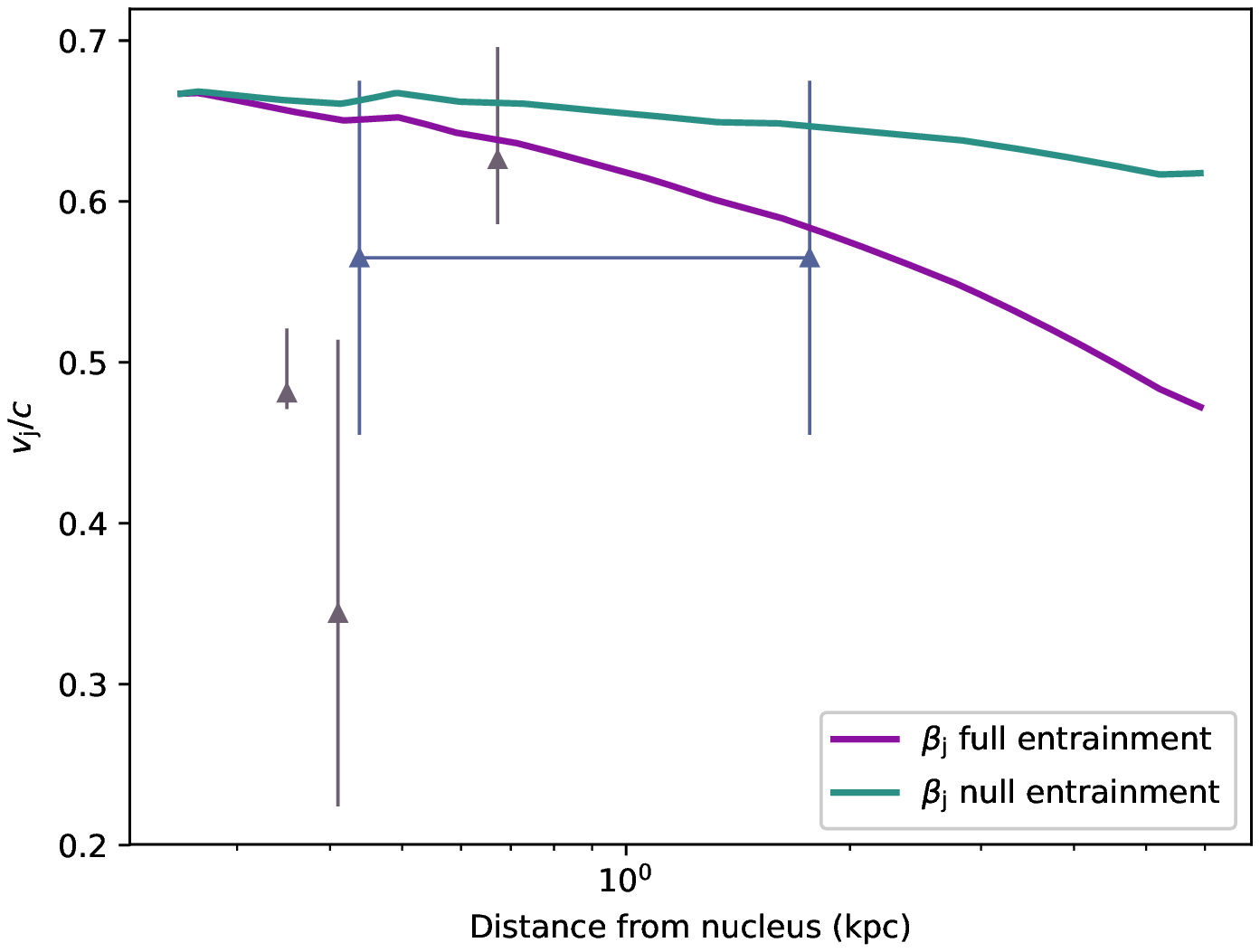}
	\end{tightcenter}
	\caption{Same as Fig.\,\ref{fig:fluid-model-fig5} but for entrained fraction of 0 (green line) over the modelled region.}
	\label{fig:fluid-model-figC1}
\end{figure}

\begin{figure}
	\begin{tightcenter}
	\includegraphics[width=0.53\textwidth]{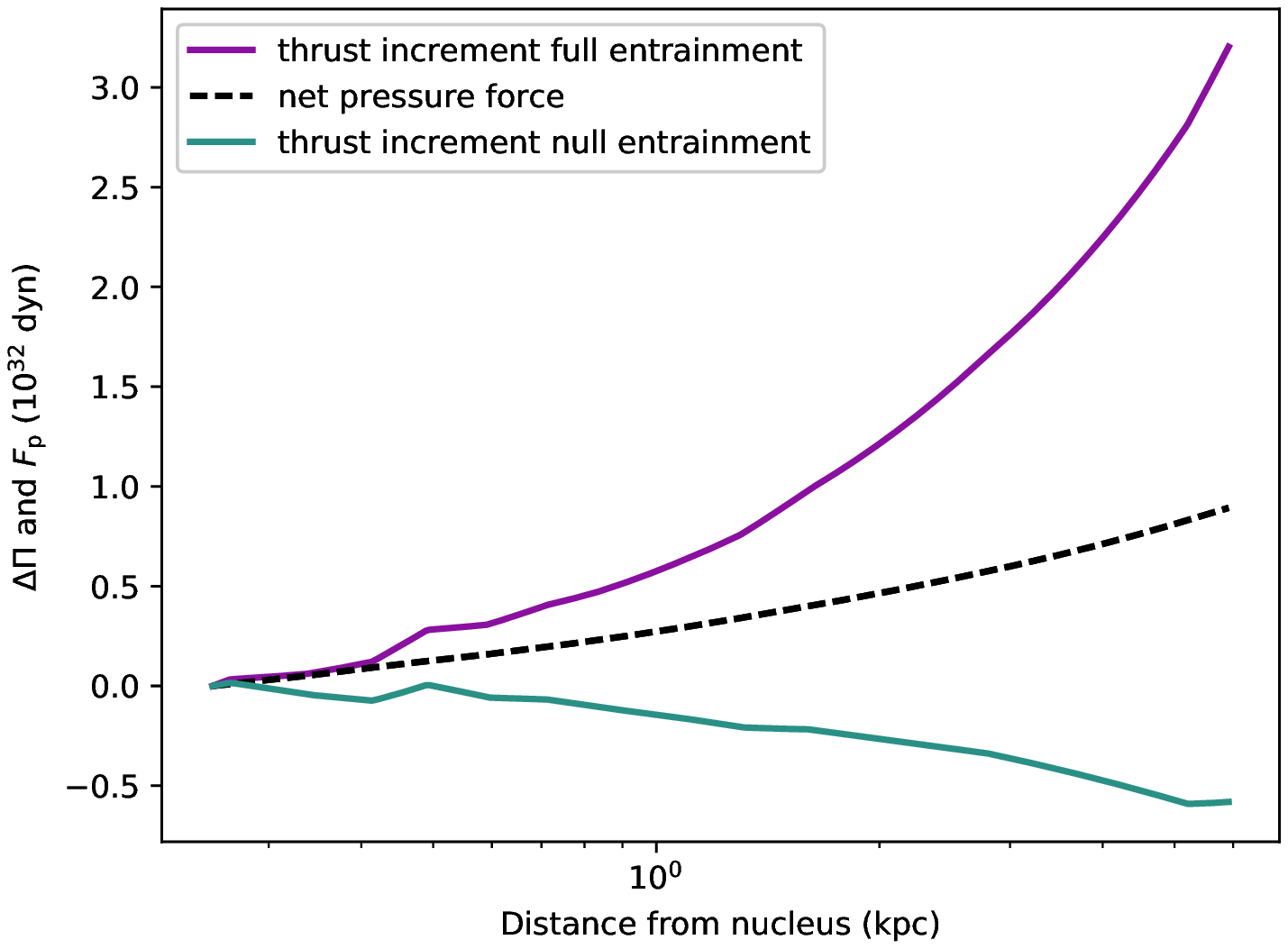}
	\end{tightcenter}
	\caption{Same as Fig.\,\ref{fig:fluid-model-fig7} but for entrained fraction of 0 (green line) over the modelled region.}
	\label{fig:fluid-model-figC2}
\end{figure}

To demonstrate the effect of zero mass entrainment, we replot the jet velocity, and the momentum flux increment and net pressure force versus distance from the nucleus, in respectively Figs\,\ref{fig:fluid-model-figC1} and \ref{fig:fluid-model-figC2}.

Fig.\,\ref{fig:fluid-model-figC1} shows that the jet velocity for zero entrainment drops slightly as the jet propagates. This could be due to external drag and the effect of the small quantity of material that is entrained `in front' of the modelled region, or to how sensitive the decrease is to our assumed parameters. 

The apparent decrease in jet momentum for zero entrainment in Fig.\,\ref{fig:fluid-model-figC2} tells us that either there is appreciable external drag on the jet, or it entrains mass. We cannot firmly rule out the former possibility, although the drag might not be large given the marginal zero-entrainment velocity drop. The X-ray knots might be sites where the jet interacts with nearly-stationary obstacles that it does not entrain, so they tap momentum from the jet -- though not appreciable energy. However, the fact that entraining the mass shed by stars produces about the right change in momentum flux indicates that our interpretation is the more likely. If the mass shed by the stars was not accelerated to speeds comparable to the jet, it would not exert nearly enough drag to account for the momentum loss; therefore, it needs to be `fully' entrained to get the adequate change in momentum flux.


\bsp	
\label{lastpage}
\end{document}